\documentclass[aps,preprint,prd,superscriptaddress,preprintnumbers,nofootinbib]{revtex4-1}[11pt]
\pdfoutput=1
\usepackage{graphicx}
\usepackage{mathrsfs}
\usepackage{slashed}
\usepackage{latexsym}
\usepackage{hyperref}
\usepackage{amsmath}
\usepackage{url}
\usepackage{appendix}
\usepackage{indentfirst}
\usepackage{color}

\begin{document}

\title{One-Loop Radiative Correction to the Triple Higgs Coupling \\ in the Higgs Singlet Model}
\date{\today}
\author{Shi-Ping He}
\email{sphe@pku.edu.cn}
\affiliation{Institute of Theoretical Physics $\&$ State Key Laboratory of Nuclear
Physics and Technology, Peking University, Beijing 100871, China}
\author{Shou-hua Zhu}
\email{shzhu@pku.edu.cn}
\affiliation{Institute of Theoretical Physics $\&$ State Key Laboratory of Nuclear
Physics and Technology, Peking University, Beijing 100871, China}
\affiliation{Center for High Energy Physics, Peking University, Beijing 100871, China}
\affiliation{Collaborative Innovation Center of Quantum Matter, Beijing 100871, China}

\begin{abstract}

Though the 125 GeV Higgs boson is consistent with the standard model (SM) prediction until now, the triple Higgs coupling can deviate
from the SM value in the physics beyond the SM (BSM). In this paper, the radiative correction to the triple Higgs coupling is investigated in the minimal extension of the SM by adding a real gauge singlet scalar. In this model there are two scalars $h$ and $H$, and both of them are mixing states of the doublet and singlet. Provided that the mixing angle is set to be zero, namely the SM limit, $h$ is the pure left-over of the doublet and its behaviour is the same as that of the SM at the tree level. However the loop corrections can alter $h$-related couplings. In this SM limit case, the effect of the singlet $H$ may show up in the $h$-related couplings, especially the triple $h$ coupling. Our numerical results show that the deviation is sizable. For $\lambda_{\Phi{S}}=1$ (see the text for definition), the deviation $\delta_{hhh}^{(1)}$ around threshold can be more than $20\%$ if mass of the new scalar is light. For $\lambda_{\Phi{S}}=1.5,3$, the deviation $\delta_{hhh}^{(1)}$ can reach above $30\%$ in the vicinity of threshold. For $\lambda_{\Phi{S}}=-1$, $m_H$ can be light. The deviation $\delta_{hhh}^{(1)}$ is negative in this case and can be even $-50\%$ in the threshold enhanced region. In the optimal case, the triple $h$ coupling can be measured to be $20\%$, thus it is very sensitive to the BSM physics. The radiative correction for the $hZZ$ coupling is from the counter-term, which is universal in this model and always with negative sign. $\delta_{hZZ}^{(1)}$ is independent of the sign of $\lambda_{\Phi{S}}$ and its typical size is $-0.2\%\sim-2\%$. The $hZZ$ coupling, which can be measured at percent and even sub-percent level at future high luminosity electron-positron colliders, may be a complementarity to the triple $h$ coupling to search for the BSM. When combining $hhh$ and $hZZ$ precision measurements together, parameter space of this model can be probed well including the sign and size of $\lambda_{\Phi{S}}$.
\end{abstract}

\maketitle

\section{Introduction}\label{sec:1}
\indent{ The} standard model (SM) has been extensively tested, especially the deviations for the gauge sector are strongly constrained by the electro-weak precision measurements from the Large Electron-{Positron} Collider (LEP) \cite{Agashe:2014kda}, Tevatron and the Large Hadron Collider (LHC). However, the Yukawa sector and the scalar sector are two sectors which are still not well probed.  Since the discovery of the Higgs boson at the LHC in 2012 \cite{Aad:2012tfa,Chatrchyan:2012xdj}, the most important task is to measure the properties of the scalar accurately. The measurements will help us understand the nature of the electro-weak symmetry breaking mechanism (EWSB) \cite{Englert:1964et,Higgs:1964ia,Guralnik:1964eu,Kibble:1967sv}. If there exists new physics beyond the SM (BSM), it is believed that it is related with the Higgs couplings more or less. The Higgs boson is a door to the unknown new world.

\indent{Current} measurements of the Higgs couplings with gauge bosons tend to be the SM values \cite{Khachatryan:2016vau, Sirunyan:2018koj}. At the same time Higgs couplings with the third generation fermions ($\tau$ lepton Yukawa \cite{Sirunyan:2017khh, Aaboud:2018pen}, top quark Yukawa \cite{Sirunyan:2018hoz, Aaboud:2018urx}, bottom quark Yukawa \cite{Sirunyan:2018kst, Aaboud:2018zhk}) are also consistent with those in the SM. Usually for the model construction, the Higgs couplings with fermions and gauge bosons will have the SM limit at the electro-weak scale. However the triple Higgs coupling can deviate from the SM value largely in this limit. Such feature of the triple Higgs coupling has been studied extensively in the two Higgs doublet model (THDM) \cite{Kanemura:2004mg,Osland:2008aw,Arhrib:2008jp}, inert Higgs doublet model (IHDM) \cite{Arhrib:2015hoa}, Higgs triplet model (HTM) \cite{Aoki:2012jj} and models with an additional heavy neutrino \cite{Baglio:2016ijw}.

\indent{Searching} for BSM physics is one of the most important goals of high energy physics. The most direct way is to increase the energy of the colliders and see whether there are new heavy resonances, while it is always hard or even impossible to construct the very high energy colliders because of the limitations from the expanses, technologies and so on. However there are other methods to achieve this goal. The new heavy particles will leave footprints at the electro-weak scale through loop effects. We may have indirect signals for the BSM through some physical quantities which are sensitive to the heavy particles.

\indent{The} minimal extension of the SM in the scalar sector is to add a real gauge singlet. The Higgs singlet model (HSM) has been studied exhaustively in a lot of papers. For example, Ref. \cite{Pruna:2013bma,Robens:2015gla} studied a model which includes a $Z_2$ symmetry spontaneously breaking real Higgs singlet and the author considered the theoretical and phenomenological constraints on this model. Ref. \cite{Chen:2014ask} explored the resonant di-Higgs production at the 14TeV hadron collider with an additional intermediate, heavy mass Higgs boson. Ref. \cite{Barger:2007im} considered two scenarios: there was (no) mixing between the SM Higgs and the singlet. Then, they analyzed the constraints from electro-weak precision observables (EWPO), LHC Higgs phenomenology and dark matter phenomenology. Ref. \cite{Falkowski:2015iwa} analyzed direct and indirect constraints on the parameter regions and the prospects for observing the decay of the heavier state into a pair of the 125 GeV Higgs. Ref. \cite{Choi:1993cv,Ham:2004cf} discussed the electro-weak phase transition (EWPT) in this model. Ref. \cite{Camargo-Molina:2016moz} calculated all one-loop scalar vertices using the effective potential approach. Ref. \cite{Bojarski:2015kra} emphasized the heavy-to-light Higgs boson decay at the electro-weak next-to-leading (NLO) order. Ref. \cite{Kanemura:2015fra} focused on the one-loop radiative corrections in the HSM and performed the numerical calculations for the $hZZ$, $hWW$, $hf\bar{f}$, $h\gamma\gamma$, $h\gamma{Z}$, $hgg$ couplings, but not for triple $h$ coupling, which is the main topic in this paper.

\indent{In} the following, we will make a careful analysis of the triple $h$ coupling up to one-loop level in this model in the SM limit. There will be an universal deviation from the SM predictions for the $hVV$, $hf\bar{f}$ couplings arising from the wave-function renormalization constant $\delta{Z_h}$. The numerical results show that the universal correction is small. For the triple $h$ coupling, there are still $hHH,hhHH$ couplings (see Appendix \ref{app:A}) in this limit. When the mass of the additional scalar is not much heavy and the coupling $\lambda_{\Phi{S}}$ is order one, the radiative correction to the triple $h$ coupling can be large in the vicinity of double Higgs production. It may be measured at future electron-positron colliders.

\indent{This} paper is organized as follows. In Sec.~\ref{sec:model}, we give a detailed description of the model and analyze the theoretical constraints on parameter space including bounded from below, perturbative unitarity and global minimum conditions. In Sec.~\ref{sec:deviation}, we firstly compute the analytic radiative correction expressions to the triple $h$ \& $hZZ$ couplings in the SM limit. Then we present the numerical results and detectability at future high energy colliders. Sec.~\ref{sec:conclusion} is devoted to the conclusions and discussions. Feynman rules, related Feynman diagrams and calculational details are collected in the Appendix.
\section{Model}\label{sec:model}
\indent{We} introduce a real additional gauge singlet $S$ with hyper-charge $Y=0$ besides the SM Higgs doublet $\Phi$. Then, we can write the scalar potential $V(\Phi,S)$ as
\begin{align}\label{eq:lag:0}
V(\Phi,S)=-m_{\Phi}^2\Phi^{\dag}\Phi+\lambda_{\Phi}(\Phi^{\dag}\Phi)^2+\mu_{\Phi{S}}\Phi^{\dag}\Phi{S}+\lambda_{\Phi{S}}
\Phi^{\dag}\Phi{S^2}+t_SS+m_S^2S^2+\mu_SS^3+\lambda_SS^4.
\end{align}
Evidently, the singlet doesn't have any Yukawa interactions or gauge interactions with the SM fields.
The scalar fields $\Phi,S$ in the unitary gauge can be parameterized as
\begin{equation}
\Phi=
\left[\begin{array}{c}
0\\h^0
\end{array}\right],h^0=\frac{v+h_1}{\sqrt{2}}
(v\approx246\mathrm{GeV}),S=v_S+h_2.
\end{equation}
Without loss of generality, we can set $v_S$ to be zero by shifting the $S$, namely redefinition of the field.
After EWSB, the two tadpoles are
$-T_{h_1}=v(\lambda_{\Phi}v^2-m_{\Phi}^2),-T_{h_2}=t_S+\frac{\mu_{\Phi{S}}}{2}v^2$. $T_{h_1},T_{h_2}$ are the coefficients in front of the fields $h_1,h_2$ in the Lagrangian.
At tree level, $T_{h_1}=0,T_{h_2}=0$, which means $m_{\Phi}^2=\lambda_{\Phi}v^2,t_S=-\frac{\mu_{\Phi{S}}}{2}v^2$.
Mass terms of the scalar fields are
\begin{equation}
\begin{aligned}
&\mathscr{L}_{mass}=-\frac{1}{2}
\left[\begin{array}{cc}
h_1&h_2\\
\end{array}\right]
\left[\begin{array}{cc}
M_{11}^2&M_{12}^2\\
M_{12}^2&M_{22}^2
\end{array}\right]
\left[\begin{array}{c}
h_1\\
h_2
\end{array}\right]\\
&M_{11}^2=2\lambda_{\Phi}v^2,M_{12}^2=\mu_{\Phi{S}}v,M_{22}^2=2m_S^2+
\lambda_{\Phi{S}}v^2.
\end{aligned}
\label{eq:lmass1}
\end{equation}
After diagonalizing the mass matrix, we get the following expressions
\begin{equation}
\begin{aligned}
&\mathscr{L}_{mass}=-\frac{1}{2}\left[\begin{array}{cc}
h&H\\
\end{array}\right]
\left[\begin{array}{cc}
m_h^2&0\\
0&m_H^2
\end{array}\right]
\left[\begin{array}{c}
h\\
H
\end{array}\right],
\left[\begin{array}{c}
h_1\\
h_2\\
\end{array}\right]=
\left[\begin{array}{cc}
\mathrm{cos}\alpha&\mathrm{sin}\alpha\\
-\mathrm{sin}\alpha&\mathrm{cos}\alpha
\end{array}\right]
\left[\begin{array}{c}
h\\
H
\end{array}\right]\\
&m_h^2=\mathrm{cos}^2\alpha{M}_{11}^2+\mathrm{sin}^2\alpha{M}_{22}^2-
\mathrm{sin}2\alpha{M}_{12}^2,~m_H^2=\mathrm{sin}^2\alpha{M}_{11}^2+\mathrm{cos}^2\alpha{M}_{22}^2
+\mathrm{sin}2\alpha{M}_{12}^2\\
&\mathrm{tan}2\alpha=\frac{2M_{12}^2}{M_{22}^2-M_{11}^2}=\frac{2\mu_{\Phi{S}}v}{2m_S^2-(2\lambda_{\Phi}-\lambda_{\Phi{S}})v^2}.
\end{aligned}
\label{eq:lmass2}
\end{equation}
In the above expressions, we use $m_H$ instead of $m_S$ to avoid the confusion with the parameter in the Lagrangian. $s_{\alpha},c_{\alpha},s_{2\alpha}$ are the simplified notations for $\mathrm{sin}\alpha,\mathrm{cos}\alpha,\mathrm{sin}2\alpha$. Here $h$ labels the SM-like Higgs boson. From now on, we will choose the parameters $m_h^2,m_H^2,\alpha,\lambda_{\Phi{S}},\lambda_S,\mu_S,v$ as the inputs.
According to the definitions of $m_h^2,m_H^2,\mathrm{tan}2\alpha$ in Eq. \eqref{eq:lmass1} and Eq. \eqref{eq:lmass2} , $\lambda_{\Phi},m_S^2,\mu_{\Phi{S}}$ can be expressed by the new inputs as
\begin{equation}
\lambda_{\Phi}=\frac{1}{2v^2}(c_{\alpha}^2m_h^2+s_{\alpha}^2m_H^2),m_S^2=\frac{c_{\alpha}^2m_H^2+s_{\alpha}^2m_h^2}{2}-\frac{1}{2}\lambda_{\Phi{S}}v^2,\mu_{\Phi{S}}=\frac{s_{2\alpha}}{2v}(m_H^2-m_h^2).
\end{equation}
Among the seven input parameters $m_h^2,m_H^2,\alpha,\lambda_{\Phi{S}},\lambda_S,\mu_S,v$ in the scalar sector, $m_h,v$ have been measured experimentally. Thus, only five free parameters $m_H^2,\alpha,\lambda_{\Phi{S}},\lambda_S,\mu_S$ are left. If we fix the value of $\alpha$ further, it becomes the four dimensional parameter space of $m_H^2,\lambda_{\Phi{S}},\lambda_S,\mu_S$.
\subsection{Constraints on the parameter space}
In the SM limit ($\alpha\rightarrow0$), $H$ will decouple from the fermions and gauge bosons because of the scaling factor $s_{\alpha}$. Thus, it will evade most of the present experimental constraints. The SM Higgs will only couple with $H$ by the interacting vertices $hHH,hhHH$. And this makes great influence on the triple $h$ coupling which is discussed later. All the analyses below will be carried out under the SM limit assumption, namely $\alpha=0~(\mu_{\Phi{S}}=0)$. Gauge interactions are the same as those in the SM, thus the model in this limit is loosely constrained from low energy to high energy experiments. Besides, the new scalar $H$ can decay through $HHH$ vertex induced loop diagram. Therefore it can't be the dark matter candidate, and there are no constraints from dark matter experiments. 

\begin{figure}[h!]
\begin{center}
\includegraphics[scale=0.4]{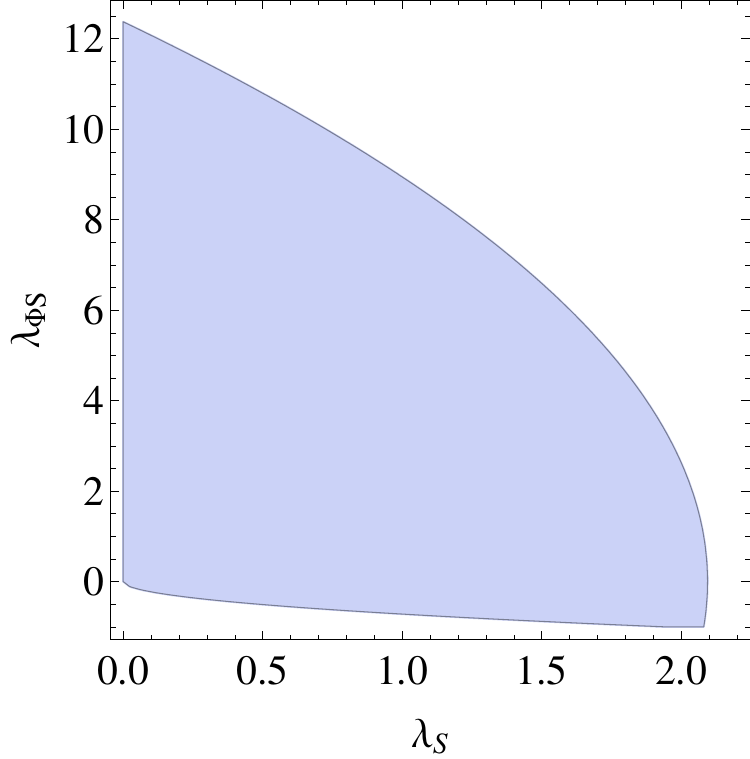}
\caption{The allowed parameter space (blue area) of $\lambda_{S},\lambda_{\Phi{S}}$ in the SM limit from Eq.~\eqref{eqn:vacuum} and~\eqref{eqn:unitarity}.}
\label{figure1}
\end{center}
\end{figure}

Due to the absence of experiment constraints, it is important to give a comprehensive analysis of the theoretical constraints. When $\Phi,S$ is very large, the scalar potential will become $V(\Phi,S)=\lambda_{\Phi}(\Phi^{\dag}\Phi)^2+\lambda_{\Phi{S}}
\Phi^{\dag}\Phi{S^2}+\lambda_SS^4$. It must be bounded from below, so we have
\begin{equation}
\lambda_{\Phi}>0,\lambda_S>0,\lambda_{\Phi{S}}>-2\sqrt{\lambda_{\Phi}\lambda_S}.
\label{eqn:vacuum}
\end{equation}
Further constraints we should consider are the so-called perturbative unitarity. S-wave amplitude $a_0$ should satisfy the relation $ |\mathrm{Re}(a_0)|<\frac{1}{2}$, where $a_0$ is given by $a_0=\frac{1}{16\pi{s}}\int_{-s}^0dt{\mathscr{M}(t)}$. Here, $s, t$ are the Mandelstam variables as usual, and $\mathscr{M}$ is the scattering amplitude. According to the Goldstone equivalence theorem, massive vector boson is dominated by the longitudinal polarization at high energy. So we only need to consider the two-to-two scattering processes with initial and final states of $W_L^+W_L^-,Z_LZ_L,Z_Lh,Z_LH,hh,HH,hH$. Similar analyses have been discussed in many papers \cite{Dawson:1988va,Cynolter:2004cq}. This is a $7\times7$ matrix, but it will be reduced into a $4\times4$ matrix in the SM limit.
A subtlety one may caution is an extra $\frac{1}{\sqrt{2}}$ for the same initial and final states, which is often ignored in many papers. After some trivial calculations (see Appendix \ref{app:unitarity}), we have the constraints from perturbative unitarity
\begin{equation}
\lambda_{\Phi}<4\pi,\lambda_{\Phi{S}}<4\pi,3\lambda_{\Phi}+6\lambda_S+
\sqrt{(3\lambda_{\Phi}-6\lambda_S)^2+4\lambda_{\Phi{S}}^2}<8\pi.
\label{eqn:unitarity}
\end{equation}
\indent{In} the SM limit, $\lambda_{\Phi}=\frac{m_h^2}{2v^2}$. Together with the bounded constraints in Eq. \eqref{eqn:vacuum}, we get the following parameter space in Fig. \ref{figure1}.
\subsection{How light can the new scalar be?}
The interesting feature is that there are no constraints on $m_H$ from Eq.~\eqref{eqn:vacuum} and~\eqref{eqn:unitarity}. Hence it is natural to ask how light can $m_H$ be. The mass of this new scalar has great influence on the test of this model. If its mass is light and the $\lambda_{\Phi{S}}$ is large, the deviations for $hhh,hZZ$ couplings may be considerable.

In the SM limit, mass formula of the new scalar can be simplified as: $m_H^2=2m_S^2+\lambda_{\Phi{S}}v^2$. Two cases may lead to a light $H$ from naive observation of this mass relation \footnote{We would like to thank Qing-Hong Cao \textit{et al} for pointing out the unreasonable parameter choice in our previous work. Where we have naively set $\lambda_{\Phi{S}}=1.5$ and $m_H\sim100\mathrm{GeV}$ at the same time without careful check, which will be examined in the following.}. The first case is $m_S^2>0,\lambda_{\Phi{S}}<0$, because there may be some cancellations betweens $m_S^2$ and $\lambda_{\Phi{S}}v^2$ to produce light $m_H$. The second case is $\lambda_{\Phi{S}}>0$ but $m_S^2<0$. You may quickly refute this case, because it can be inconsistent with the $v_S=0$ choice. In the SM, the Higgs does acquire the vacuum expectation value because the mass squared term is negative. But things are quite different from SM here, $v_S=0$ is used to reduce one redundant parameter and simplify our analysis, which is just the consequence of potential form shift invariance. Retrospecting EWSB in the SM carefully, you will get the key point. The reason is that the potential of SM at $\langle h^0\rangle=v/\sqrt{2}$ is deeper than that one at $\langle h^0\rangle=0$ and this is the most basic judgement. Similarly HSM vacuum should not only be the local minimum, but also must be a global minimum. Similar considerations have been taken into in \cite{Espinosa:2011ax, Chen:2014ask}.

We can re-parametrize $V(\Phi,S)$ as follows by eliminating $m_{\Phi}^2,t_S$ in the original potential \footnote{The constant term can be dropped.}
\begin{align}
&V(\Phi,S)=\lambda_{\Phi}(\Phi^{\dag}\Phi-\frac{v^2}{2})^2+\mu_{\Phi{S}}(\Phi^{\dag}\Phi-\frac{v^2}{2}){S}+\lambda_{\Phi{S}}
(\Phi^{\dag}\Phi-\frac{v^2}{2}){S^2}\nonumber\\
&+(m_S^2+\frac{\lambda_{\Phi{S}}}{2}v^2)S^2+\mu_SS^3+\lambda_SS^4.\nonumber\\
\end{align}
In Eq.~\eqref{eq:lag:0}, we take $(v,0)$ as the local minimum. Here we will check whether it is a global minimum. Next what we should do is to find other possible vacua ($\hat{v},\hat{v}_S$), and compare their potential values. Now let's replace scalar fields $\Phi,S$ with the parametrizations of
\begin{equation}
\Phi=
\left[\begin{array}{c}
0\\h^0
\end{array}\right],h^0=\frac{\hat{v}+h_1}{\sqrt{2}},S=\hat{v}_S+h_2.
\end{equation}
Stationary conditions $\frac{\partial V}{\partial h_1}|_{(h_1=0,h_2=0)}=\frac{\partial V}{\partial h_2}|_{(h_1=0,h_2=0)}=0$ of the scalar potential lead to the following two equations
\begin{align}
&\hat{v}[\lambda_{\Phi}(\hat{v}^2-v^2)+\mu_{\Phi{S}}\hat{v}_S+\lambda_{\Phi{S}}
\hat{v}_S^2]=0\nonumber\\
&\frac{\mu_{\Phi{S}}}{2}(\hat{v}^2-v^2)+\lambda_{\Phi{S}}\hat{v}^2
\hat{v}_S+2m_S^2\hat{v}_S+3\mu_S\hat{v}_S^2+4\lambda_S\hat{v}_S^3=0.
\label{eqn:pot:sta:re}
\end{align}
Then we need to solve the above two cubic equations of $\hat{v},\hat{v}_S$. There are three roots for $\hat{v},\hat{v}_S$ individually, thus $3\times3=9$ vacua may exist at most. Two cases are included depending on whether $\hat{v}\neq0$ or $\hat{v}=0$, which will be discussed in the following.

{\large{\textcircled{\small{1}}}} For $\hat{v}\neq0$, equations~\eqref{eqn:pot:sta:re} are equivalent to
\begin{align}
&\lambda_{\Phi}(\hat{v}^2-v^2)+\mu_{\Phi{S}}\hat{v}_S+\lambda_{\Phi{S}}
\hat{v}_S^2=0\nonumber\\
&\frac{\mu_{\Phi{S}}}{2}(\hat{v}^2-v^2)+\lambda_{\Phi{S}}\hat{v}^2
\hat{v}_S+2m_S^2\hat{v}_S+3\mu_S\hat{v}_S^2+4\lambda_S\hat{v}_S^3=0.
\end{align}
The equation satisfied by $\hat{v}_S$ is
\begin{align}
&\hat{v}_S[-\frac{\mu_{\Phi{S}}^2}{2\lambda_{\Phi}}+\lambda_{\Phi{S}}v^2+2m_S^2+(3\mu_S-\frac{3\mu_{\Phi{S}}\lambda_{\Phi{S}}}{2\lambda_{\Phi}})\hat{v}_S+(4\lambda_S-\frac{\lambda_{\Phi{S}}^2}{\lambda_{\Phi}})\hat{v}_S^2]=0.
\end{align}
$\hat{v}_S=0$ is one solution for this equation. In this case, we have $\hat{v}^2=v^2$. In the case of $\hat{v}_S\neq0$, there are three possibilities.

$\bullet$ If $4\lambda_S\lambda_{\Phi}=\lambda_{\Phi{S}}^2$ and $2\lambda_{\Phi}\mu_S=\lambda_{\Phi{S}}\mu_{\Phi{S}}$, there are no other roots.

$\bullet$ If $4\lambda_S\lambda_{\Phi}=\lambda_{\Phi{S}}^2$ but $2\lambda_{\Phi}\mu_S\neq\lambda_{\Phi{S}}\mu_{\Phi{S}}$, there is only one root with $\hat{v}_S=\frac{m_h^2m_H^2}{3v^2(\lambda_{\Phi{S}}\mu_{\Phi{S}}-2\lambda_{\Phi}\mu_S)}$
Where we have used the relation
$m_h^2m_H^2=M_{11}^2M_{22}^2-M_{12}^4=2\lambda_{\Phi}v^2(2m_S^2+
\lambda_{\Phi{S}}v^2)-\mu_{\Phi{S}}^2v^2$.

$\bullet$ If $4\lambda_S\lambda_{\Phi}\neq\lambda_{\Phi{S}}^2$, it is a quadratic equation. Let's define $\Delta=9v^2(2\lambda_{\Phi}\mu_S-\lambda_{\Phi{S}}\mu_{\Phi{S}})^2
-8m_h^2m_H^2(4\lambda_{\Phi}\lambda_S-\lambda_{\Phi{S}}^2)$ firstly. For the case of $\Delta\geq0$, the other two real roots for $\hat{v}_S$ are
\begin{align}
&v_S^{\pm}=\frac{3v(\lambda_{\Phi{S}}\mu_{\Phi{S}}-2\lambda_{\Phi}\mu_S)\pm
\sqrt{\Delta}}{4v(4\lambda_{\Phi}\lambda_S-\lambda_{\Phi{S}}^2)}.
\end{align}
For the case of $\Delta<0$, there are no other real roots for $\hat{v}_S$. Then we can easily get the roots of $\hat{v}^2$ from $v_S^{\pm}$, which are marked as
\begin{align}
&(v^{\pm})^2=v^2-\frac{\mu_{\Phi{S}}v_S^{\pm}+\lambda_{\Phi{S}}
(v_S^\pm)^2}{\lambda_{\Phi}}.
\end{align}

Noticing that there are six possible solutions with $(\hat{v},\hat{v}_S)=(\pm v,0),(\pm v^+,v_S^+),(\pm v^-,v_S^-)$ generally, but $\hat{v}$ and $-\hat{v}$ are symmetric. Thus six vacua are reduced into three independent vacua in the case of $\hat{v}\neq0$.

{\large{\textcircled{\small{2}}}} For $\hat{v}=0$, we have $-\frac{\mu_{\Phi{S}}}{2}v^2+2m_S^2\hat{v}_S+3\mu_S\hat{v}_S^2+4\lambda_S\hat{v}_S^3=0$ from equations~\eqref{eqn:pot:sta:re}. The three roots of this equation are labelled as $v_S^{(1)},v_S^{(2)},v_S^{(3)}$. Let's define the variables $\kappa,\Delta_0$ as
\begin{align}
&\Delta_0=-32[2m_S^4(9\mu_S^2-32\lambda_{S}m_S^2)
+27\mu_{\Phi{S}}\mu_{S}v^2(\mu_{S}^2-4\lambda_{S}m_S^2)
-54\lambda_S^2\mu_{\Phi{S}}^2v^4]\nonumber\\
&\kappa=12[-18\mu_S(\mu_S^2-4\lambda_{S}m_S^2)
+72\lambda_{S}^2\mu_{\Phi{S}}v^2+\lambda_S\sqrt{3\Delta_0}].
\end{align}
Then the explicit expressions of $v_S^{(1)},v_S^{(2)},v_S^{(3)}$ are given in the following
\begin{align}
&v_S^{(1)}=\frac{(6\mu_{S}-\kappa^{1/3})^2-96\lambda_{S}m_S^2}{24\lambda_{S}\kappa^{1/3}}+\frac{\mu_{S}}{4\lambda_{S}}\nonumber\\
&v_S^{(2)}=\frac{(6\mu_{S}-e^{2i\pi/3}\kappa^{1/3})^2-96\lambda_{S}m_S^2}{24e^{2i\pi/3}\lambda_{S}\kappa^{1/3}}+\frac{\mu_{S}}{4\lambda_{S}}\nonumber\\
&v_S^{(3)}=\frac{(6\mu_{S}-e^{4i\pi/3}\kappa^{1/3})^2-96\lambda_{S}m_S^2}{24e^{4i\pi/3}\lambda_{S}\kappa^{1/3}}+\frac{\mu_{S}}{4\lambda_{S}}.
\end{align}
Where $v_S^{(1)}$ is real and $v_S^{(2)},v_S^{(3)}$ are complex conjugates if $\Delta_0>0$. All of the roots $v_S^{(1)},v_S^{(2)},v_S^{(3)}$ are real if $\Delta_0\leq0$.

As a matter of fact, the cubic equation of $\hat{v}_S$ can be simplified as $\hat{v}_S[2m_S^2+3\mu_S\hat{v}_S+4\lambda_S\hat{v}_S^2]=0$ in the SM limit. It is easy to get the explicit expressions of $v_S^{(1)},v_S^{(2)},v_S^{(3)}$\footnote{Here $v_S^{(1)},v_S^{(2)},v_S^{(3)}$ may not correspond to above three solutions one-by-one.}
\begin{align}
&v_S^{(1)}=0\nonumber\\
&v_S^{(2)}=\frac{-3\mu_{S}+\sqrt{9\mu_{S}^2-32\lambda_{S}m_S^2}}{8\lambda_{S}}\nonumber\\
&v_S^{(3)}=\frac{-3\mu_{S}-\sqrt{9\mu_{S}^2-32\lambda_{S}m_S^2}}{8\lambda_{S}}.
\end{align}

It is time to give a summary. In the most general case, there are totally six possible nonequivalent vacua marked as
$(0,v_S^{(1)}),(0,v_S^{(2)}),(0,v_S^{(3)}),(v,0),(v_+,v_S^+),(v_-,v_S^-)$. It will lead to the condition $V(\frac{v}{\sqrt{2}},0)\leq \mathrm{Min}\{V(\frac{\hat{v}}{\sqrt{2}},\hat{v}_S)\}$ by taking $(v,0)$ as the global minimum. Here we have $(\hat{v},\hat{v}_S)\in\{(v_+,v_S^+),(v_-,v_S^-),(0,v_S^{(1)}),(0,v_S^{(2)}),(0,v_S^{(3)})\}$ and only need to consider those cases of real roots.

It is difficult to find exact solutions of the global minimum conditions, because quartic equations of $m_H^2$ are involved. But we can get the behaviour of $m_H^2$ minimum in some limits:

$\bullet$ If $-2\sqrt{\lambda_{\Phi}\lambda_S}<\lambda_{\Phi S}<2\sqrt{\lambda_{\Phi}\lambda_S}$, the conservative minimum of $m_H^2$ is zero for $\mu_S^2\rightarrow0$, while it is $\frac{2\lambda_{\Phi}\mu_S^2}{4\lambda_{\Phi}\lambda_{S}-\lambda_{\Phi S}^2}$ for $\frac{\mu_S^2}{\lambda_S}\gg \lambda_{\Phi}v^2,\lambda_Sv^2,\lambda_{\Phi S}v^2$.

$\bullet$ If $\lambda_{\Phi S}>2\sqrt{\lambda_{\Phi}\lambda_S}$, the conservative minimum of $m_H^2$ is $(\lambda_{\Phi S}-2\sqrt{\lambda_{\Phi}\lambda_S})v^2$ for $\mu_S^2\rightarrow0$, while it is $\frac{\mu_S^2}{2\lambda_S}$ for $\frac{\mu_S^2}{\lambda_S}\gg \lambda_{\Phi}v^2,\lambda_Sv^2,\lambda_{\Phi S}v^2$.

Now let's plot the parameter space region allowed by bounded from below, perturbative unitarity and global minimum constraints. When SM limit $\alpha=0$ is adopted, the allowed parameter space is symmetric under $\mu_S\rightarrow-\mu_S$. The reason is that scalar potential is invariant under the transformation of $\mu_S\rightarrow-\mu_S,S\rightarrow-S$. It is obvious that perturbativity unitarity and local minimum bounds give no constraints on $\mu_S$. What's more, the global minimum conditions depend only on $\hat{v}_S^2,\mu_S\hat{v}_S$, which are $\mu_S\rightarrow-\mu_S$ invariant. So we only need to take the positive $\mu_S$ cases into account. Then, we choose the following five cases:

$\bullet$ $\lambda_{\Phi{S}}=1,\mu_S=20,200\mathrm{GeV}$

$\bullet$ $\lambda_{\Phi{S}}=1.5,\mu_S=20,200\mathrm{GeV}$

$\bullet$ $\lambda_{\Phi{S}}=3,\mu_S=20,200\mathrm{GeV}$

$\bullet$ $\lambda_{\Phi{S}}=-0.5,\mu_S=20,200\mathrm{GeV}$

$\bullet$ $\lambda_{\Phi{S}}=-1,\mu_S=20\mathrm{GeV}$.

\begin{figure}
\begin{center}
\includegraphics[scale=0.3]{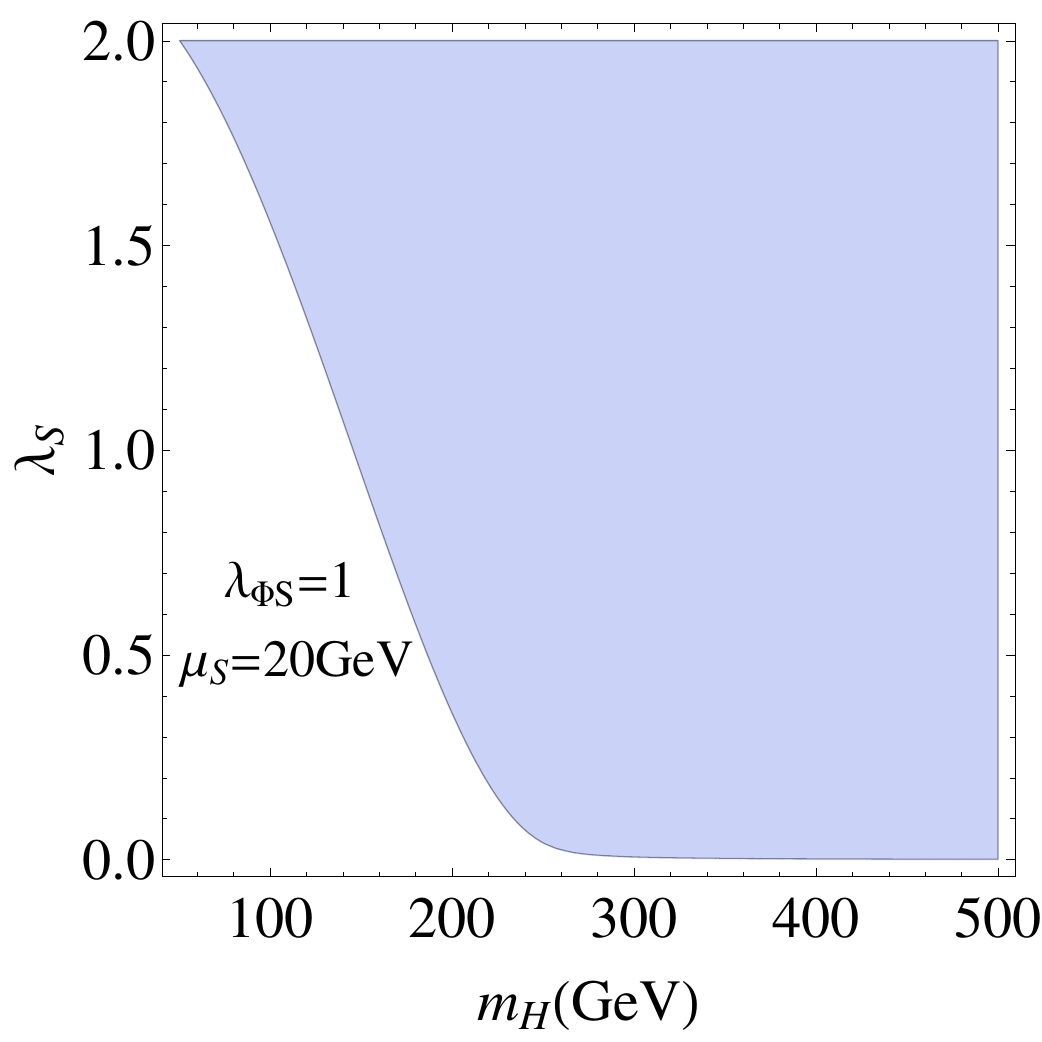}
\includegraphics[scale=0.3]{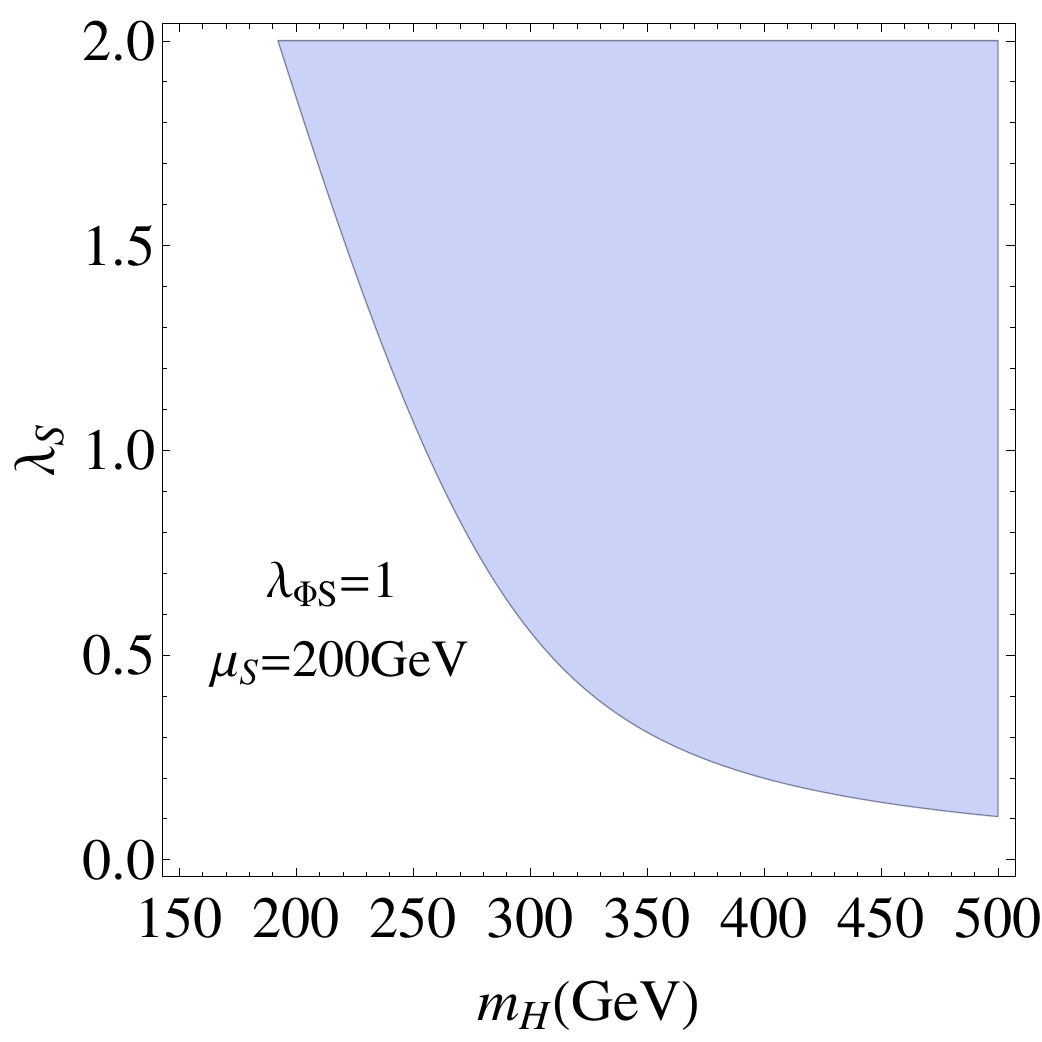}
\caption{The allowed parameter space region in $m_H,\lambda_S$ plane for $\lambda_{\Phi{S}}=1$ and $\mu_S=20~(\mathrm{left}),~200~(\mathrm{right})~\mathrm{GeV}$ respectively.}
\label{fig:para:1}
\end{center}
\end{figure}

\begin{figure}
\begin{center}
\includegraphics[scale=0.3]{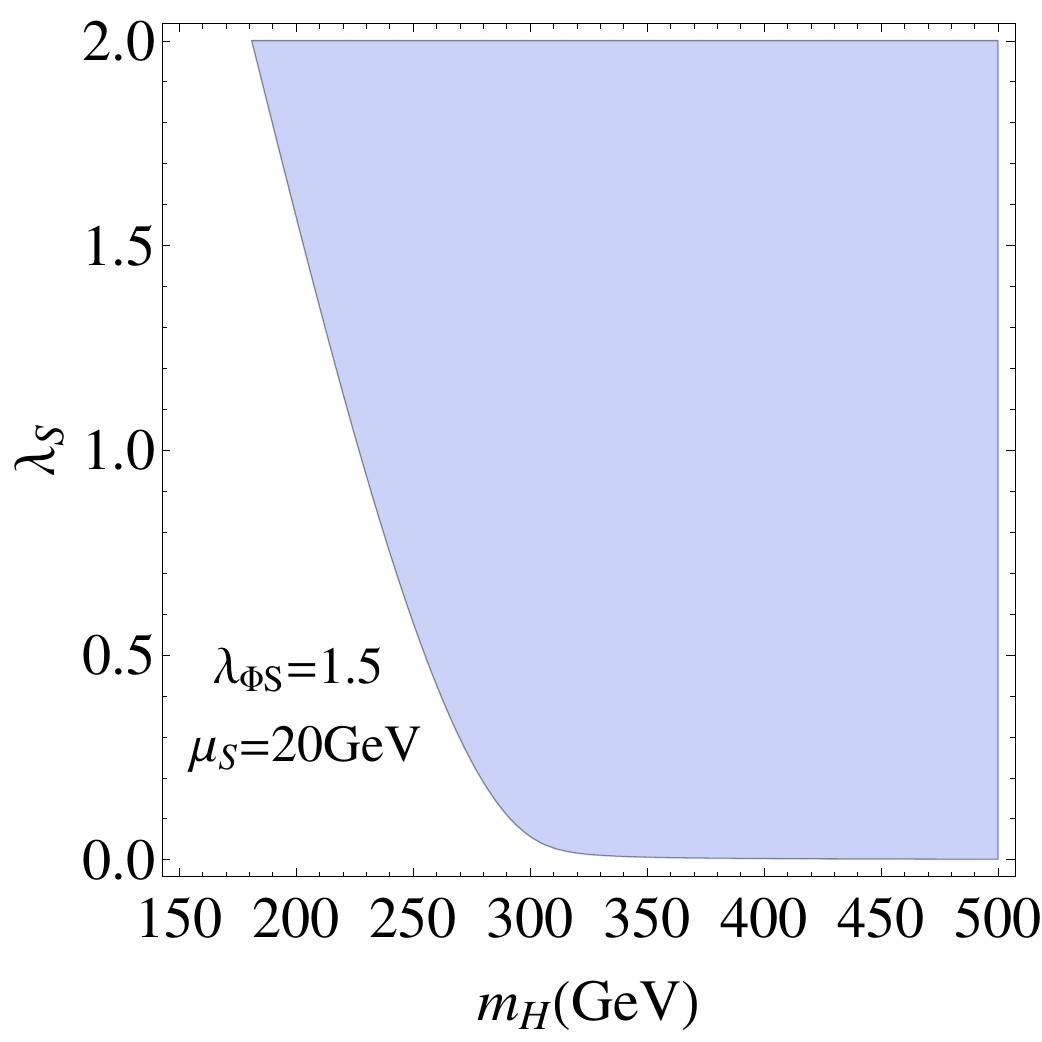}
\includegraphics[scale=0.3]{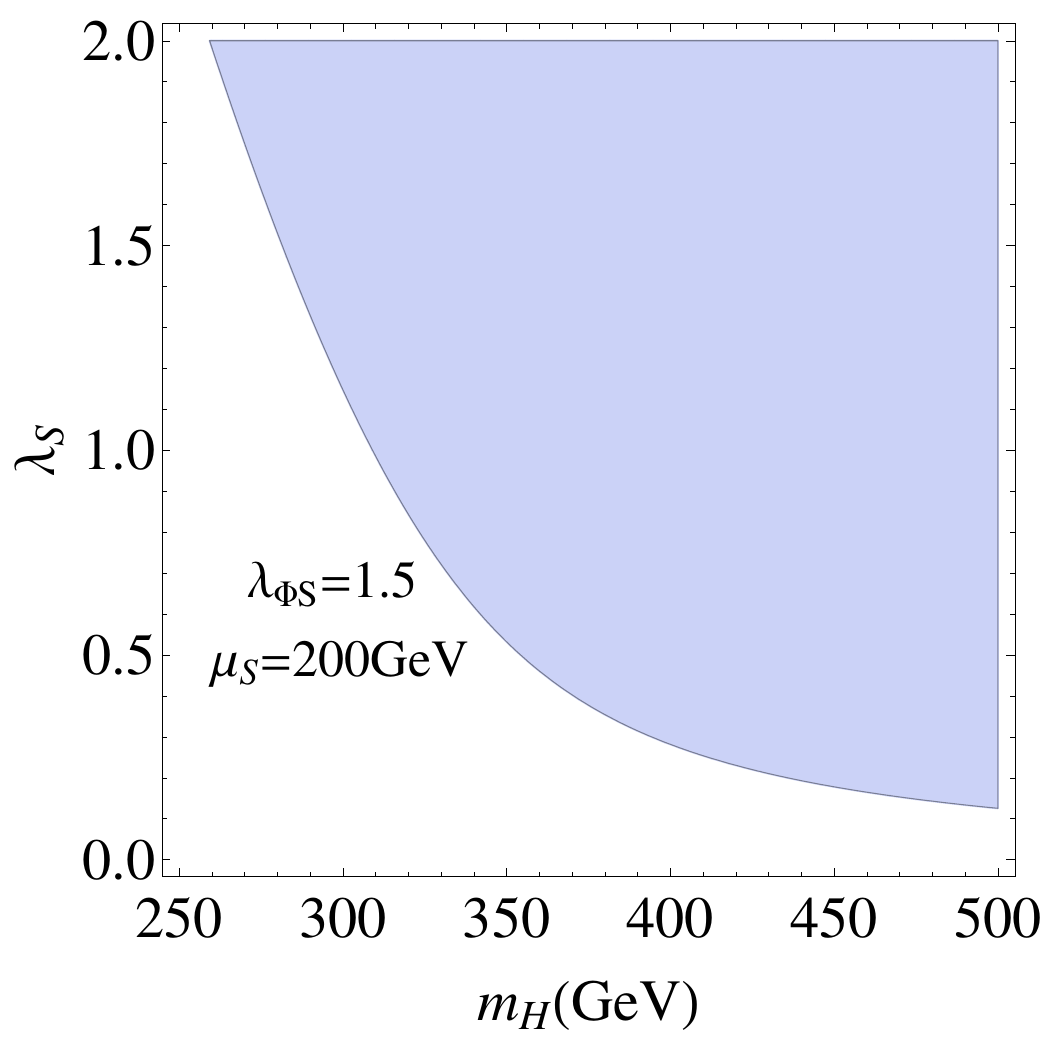}
\caption{The allowed parameter space region in $m_H,\lambda_S$ plane for $\lambda_{\Phi{S}}=1.5$ and $\mu_S=20~(\mathrm{left}),~200~(\mathrm{right})~\mathrm{GeV}$ respectively.}
\label{fig:para:1d5}
\end{center}
\end{figure}

\begin{figure}
\begin{center}
\includegraphics[scale=0.3]{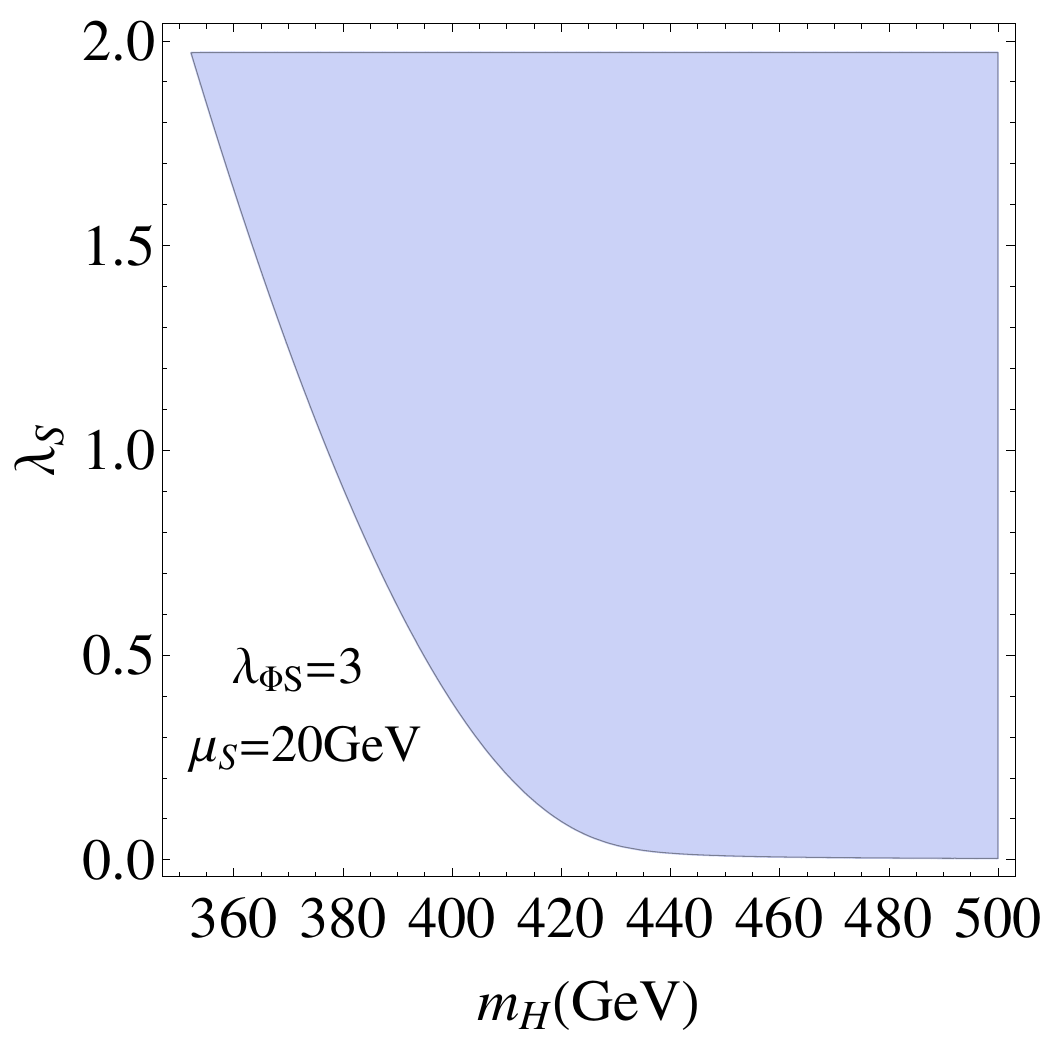}
\includegraphics[scale=0.3]{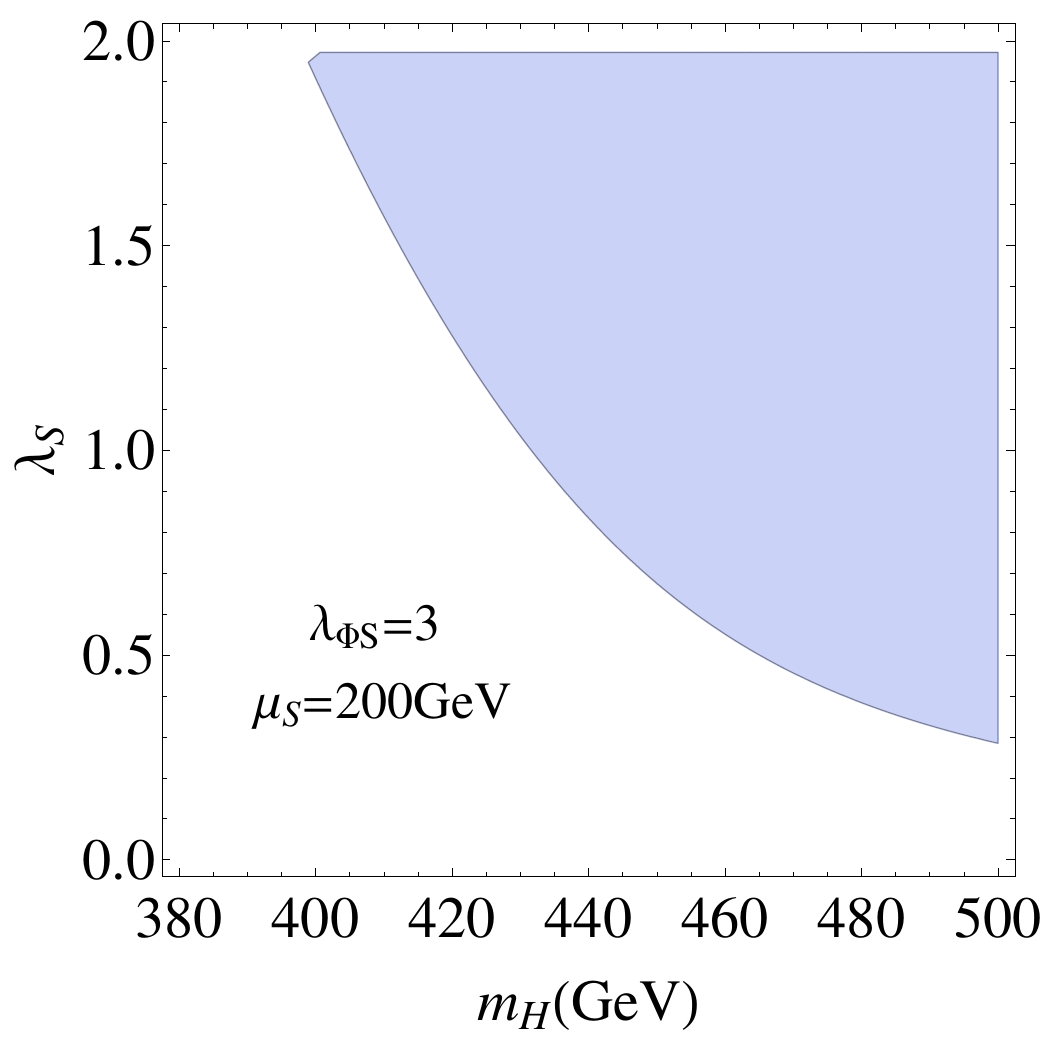}
\caption{The allowed parameter space region in $m_H,\lambda_S$ plane for $\lambda_{\Phi{S}}=3$ and $\mu_S=20~(\mathrm{left}),~200~(\mathrm{right})~\mathrm{GeV}$ respectively.}
\label{fig:para:3}
\end{center}
\end{figure}

\begin{figure}
\begin{center}
\includegraphics[scale=0.3]{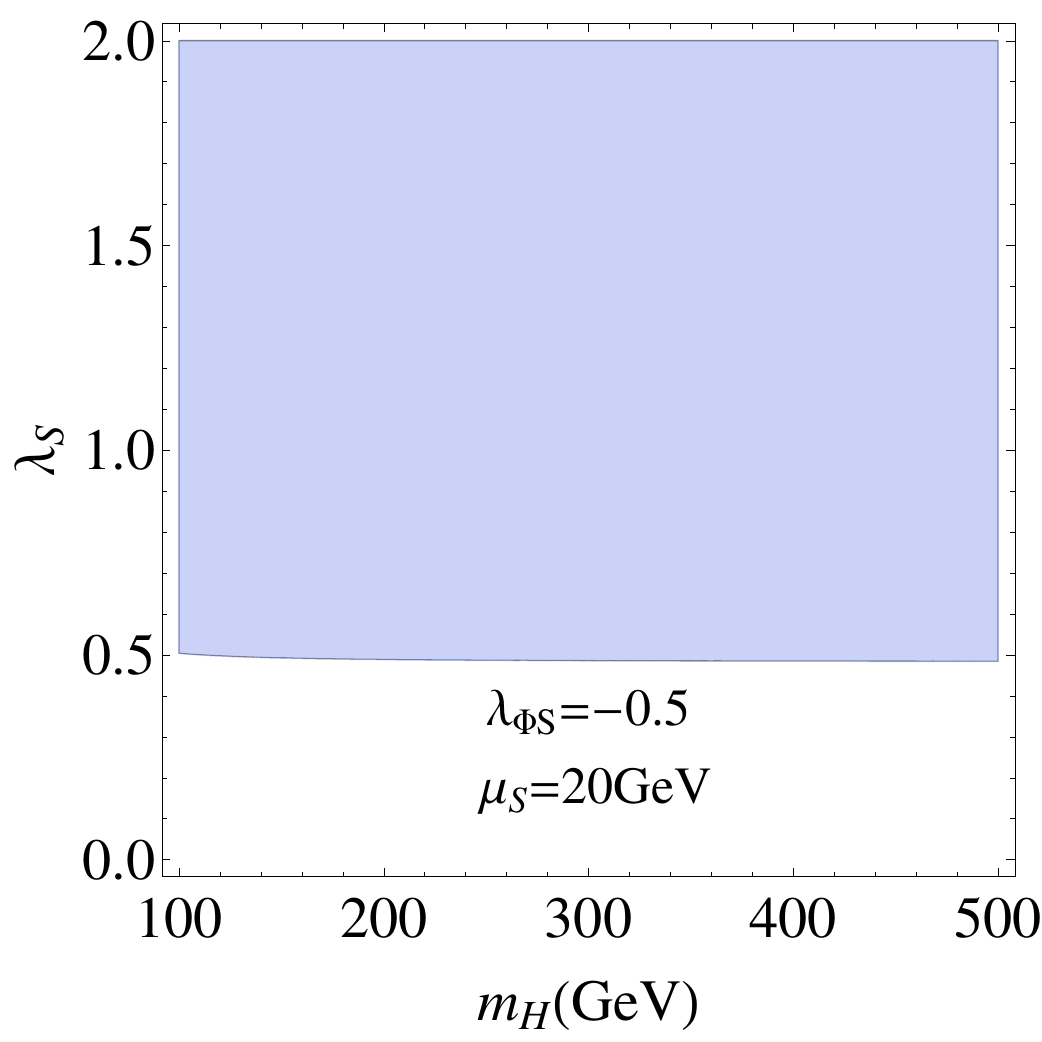}
\includegraphics[scale=0.3]{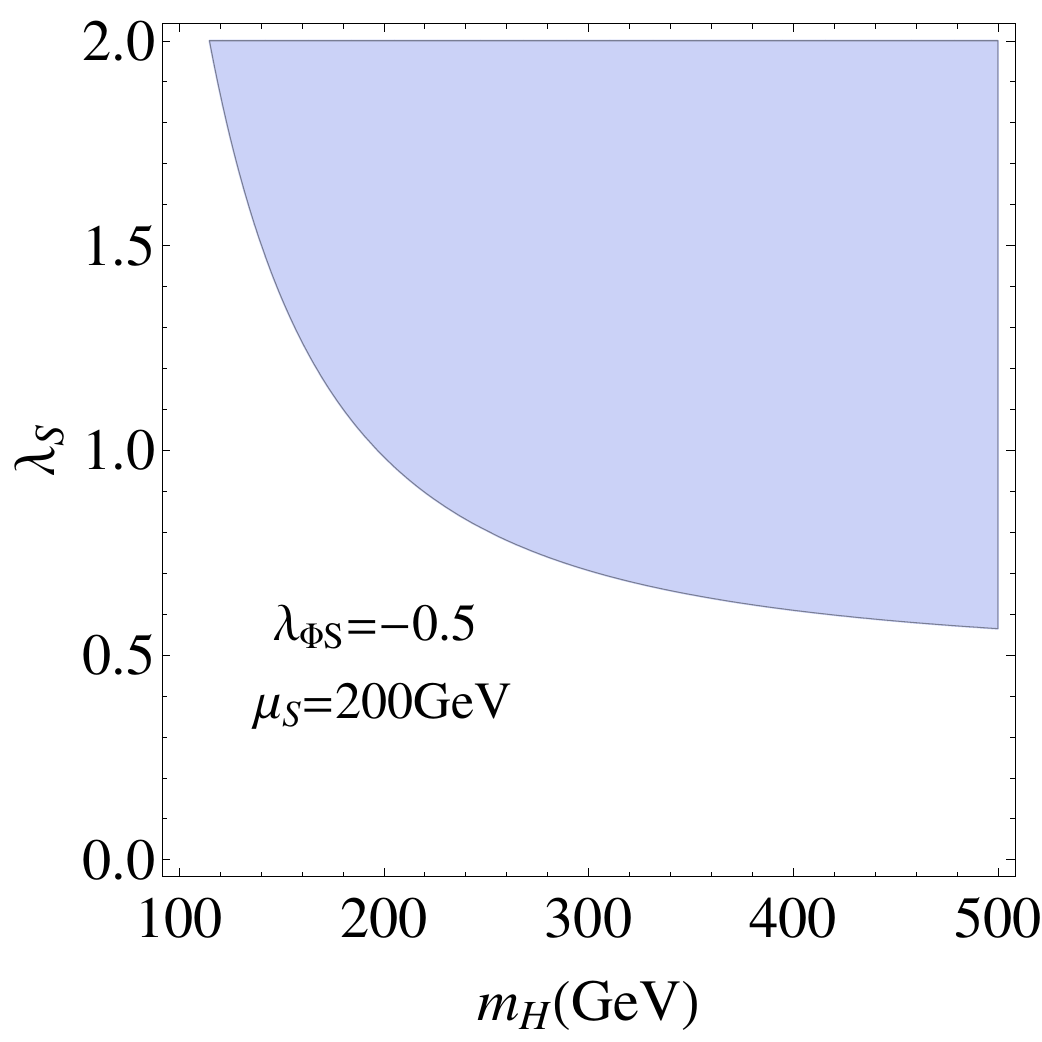}
\includegraphics[scale=0.3]{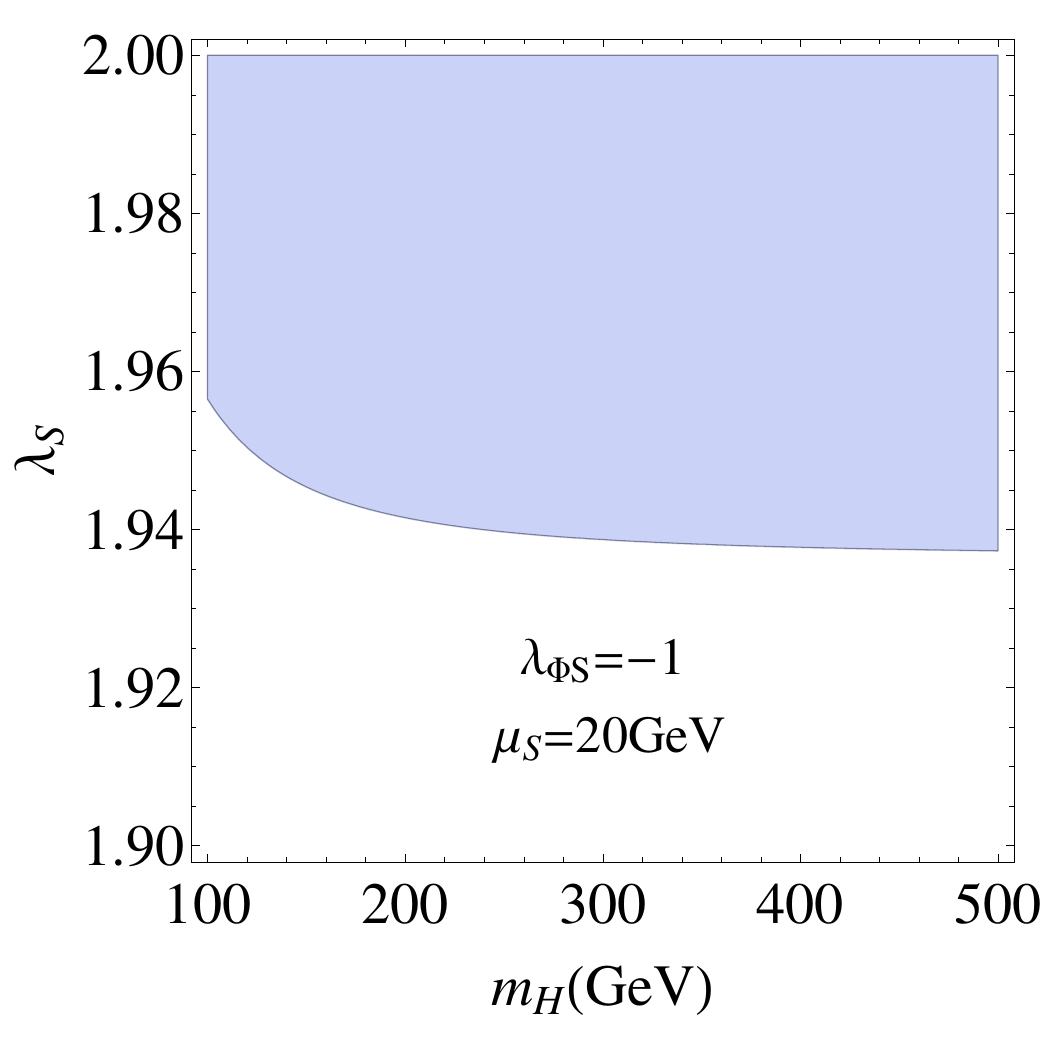}
\caption{The allowed parameter space region in $m_H,\lambda_S$ plane for $\lambda_{\Phi{S}}=-0.5$ and $\mu_S=20~(\mathrm{left}),~200~(\mathrm{middle})~\mathrm{GeV}$ respectively. In the right panel, we also plot the allowed parameter space region in $m_H,\lambda_S$ plane for $\lambda_{\Phi{S}}=-1$ and $\mu_S=20\mathrm{GeV}$. Here $\lambda_{\Phi{S}}=-1,\mu_S=200\mathrm{GeV}$ is not favoured any more for $m_H$ less than $500\mathrm{GeV}$.}
\label{fig:para:m0d5}
\end{center}
\end{figure}

In Fig~\ref{fig:para:1}~\ref{fig:para:1d5}~\ref{fig:para:3}~\ref{fig:para:m0d5}, we plot the allowed $m_H,\lambda_S$ parameter space for the above cases. We find that the minimal lower limit for $m_H$ is $80,180,350\mathrm{GeV}$ roughly in the case of $\lambda_{\Phi S}=1,1.5,3$ respectively. The larger $|\mu_S|$ becomes, the larger lower limit of $m_H$ is. When $m_H$ is light enough, the $h\rightarrow HH$ decay channel can be open. Thus Higgs invisible decay will give strong constraints. The most stringent upper limit on $\mathrm{Br}(h\rightarrow\mathrm{invisible})$ is $19\%$ at $95\%$ confidence level presently \cite{Sirunyan:2018owy}. For $m_H=50\mathrm{GeV}$, it gives the bound $|\lambda_{\Phi{S}}|<0.01$. Hence $m_H$ should be large than $\frac{1}{2}m_h$ in the case of $\lambda_{\Phi S}=-1,-0.5$.

Here are some conclusions and comments about the allowed parameter space in the SM limit:

$\bullet$ If $|\lambda_{\Phi{S}}|\geq0.01$, $m_H$ should lie above $\frac{1}{2}m_h$ because of the strong constraints from Higgs invisible decay.

$\bullet$ Bounded from below and perturbative unitarity conditions can only constrain $\lambda_S,\lambda_{\Phi{S}}$. In order to restrict $m_H,\mu_S$, global minimum conditions should be considered.

$\bullet$ The allowed parameter space is symmetric under $\mu_S\rightarrow-\mu_S$. For larger $|\mu_S|$, the global minimum conditions favour larger $m_H$. For example, $m_H$ is not allowed to be less than $500\mathrm{GeV}$ in the case of $\lambda_{\Phi{S}}=-1,\mu_S=200\mathrm{GeV}$. While it is still alive with $m_H$ less than $500\mathrm{GeV}$ for $\lambda_{\Phi{S}}=-1,\mu_S=20\mathrm{GeV}$.

$\bullet$ In the mass formula $m_H^2=2m_S^2+\lambda_{\Phi{S}}v^2$, $m_S^2$ can be negative. If $\lambda_{\Phi{S}}\sim1$, $m_H$ will have a lower limit depending on the choice of $\mu_S$. If $\lambda_{\Phi{S}}<0$, $m_H$ will be loosely constrained, but $\lambda_S$ will be strongly bounded.

$\bullet$ If $\lambda_{\Phi{S}}$ is too large, it may run into the Landau pole quickly. If $\lambda_{\Phi{S}}$ lies near -1, it may run out of the allowed region after radiative corrections. One-loop running of the parameters are quite complex and systematic discussions are beyond the ability of this work.
\section{Analytic and numerical results of the triple $h$ $\&$ $hZZ$ coupling deviations}\label{sec:deviation}
\subsection{One-loop radiative corrections to the triple $h$ $\&$ $hZZ$ couplings in the SM limit}

We will calculate the deviation of the triple $h$ coupling from the SM value originated from one-loop radiative correction in the SM limit. During the calculations, we adopt the conventions from Ref. \cite{Denner:1991kt}. There is no doubt that the loop particles must be the additional scalar $H$. To gauge the deviation from the SM value, we define $\delta_{hhh}^{(1)}$ as
\begin{equation}
\delta_{hhh}^{(1)}\equiv\frac{\lambda_{hhh}^{(\mathrm{HSM})}-
\lambda_{hhh}^{(\mathrm{SM})}}{\lambda_{hhh}^{(\mathrm{SM},tree)}}.
\label{dev1}
\end{equation}
In the following, we will present the numerical results for $\delta_{hhh}^{(1)}$ for the chosen
model parameters.

Similarly, the deviation of the $hZZ$ coupling from the SM value originated from one-loop radiative correction in the SM limit is defined as
\begin{equation}
\delta_{hZZ}^{(1)}\equiv\frac{\lambda_{hZZ}^{(\mathrm{HSM})}-
\lambda_{hZZ}^{(\mathrm{SM})}}{\lambda_{hZZ}^{(\mathrm{SM},tree)}}.
\label{dev2}
\end{equation}

Tadpole renormalization constant $\delta{t}$ is determined from vanishing of the $h$ tadpole up to one-loop. Renormalization constants $\delta{m_h^2}~,~\delta{Z_h}$ are given by the self-energy correction of the SM Higgs $h$ (see Appendix \ref{app:ren}). The analytical expressions can be found in Appendix \ref{app:triple}, where on-shell renormalization scheme is adopted.
\subsection{Numerical results}
In this section, we will do some numerical evaluations of $\delta_{hhh}^{(1)},\delta_{hZZ}^{(1)}$ for different model parameters. We set $m_h=125\mathrm{GeV},v=246\mathrm{GeV}$ as in the SM. The other parameters are chosen to maximize $\delta_{hhh}^{(1)},\delta_{hZZ}^{(1)}$ if possible. The deviation $\delta_{hhh}^{(1)}$ is mainly determined by $\lambda_{\Phi{S}},m_H,\sqrt{p^2}$, where one of the Higgs bosons with momentum $p$ is off-shell. The dominant contribution is from the triangle diagram which is proportional to $\lambda_{\Phi{S}}^3$. The deviation $\delta_{hZZ}^{(1)}$ is determined by $\lambda_{\Phi{S}},m_H$. It is originated from wave-function renormalization constant and is proportional to $\lambda_{\Phi{S}}^2$. We choose the value of $\lambda_{\Phi{S}}=1,1.5,3,-1$ with the allowed $m_H$ mass range respectively, and study the dependence on $m_H,\sqrt{p^2}$. Now we need to perform numerical analyses of the defined deviations $\delta_{hhh}^{(1)},\delta_{hZZ}^{(1)}$ using LoopTools \cite{Hahn:1998yk}. In the following, we will choose the four benchmark scenarios:
 
$\bullet$ For $\lambda_{\Phi{S}}=1$, we choose $m_H\in[80,200]\mathrm{GeV}$

$\bullet$ For $\lambda_{\Phi{S}}=1.5$, we choose $m_H\in[180,300]\mathrm{GeV}$

$\bullet$ For $\lambda_{\Phi{S}}=3$, we choose $m_H\in[350,500]\mathrm{GeV}$

$\bullet$ For $\lambda_{\Phi{S}}=-1$, we choose $m_H\in[80,180]\mathrm{GeV}$.

\begin{figure}[h]
\begin{center}
\includegraphics[scale=0.4]{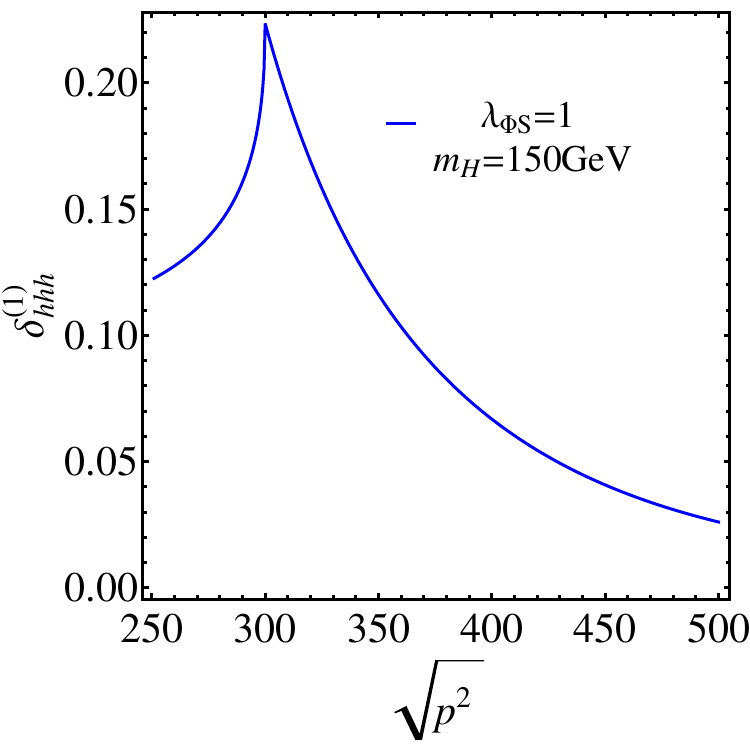}
\includegraphics[scale=0.4]{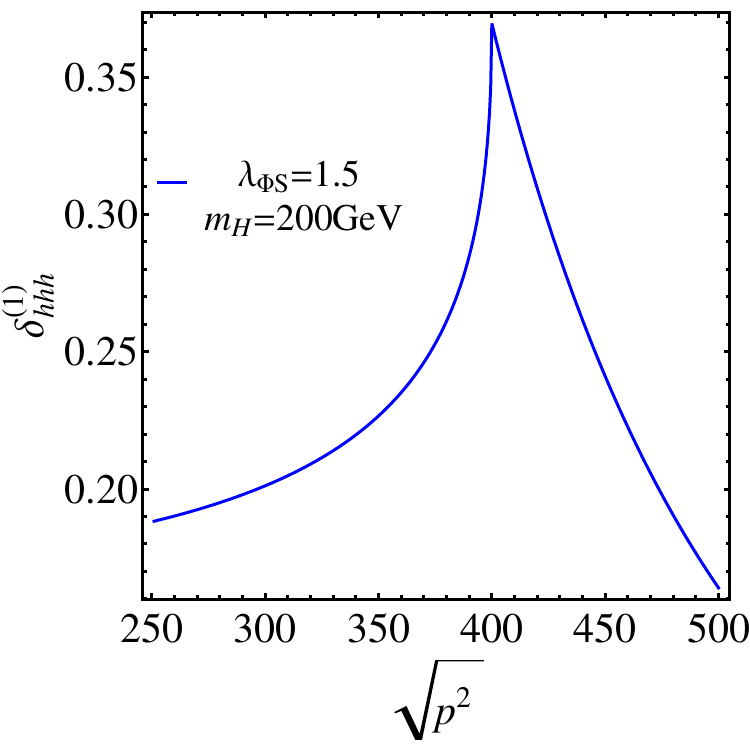}\\
\includegraphics[scale=0.4]{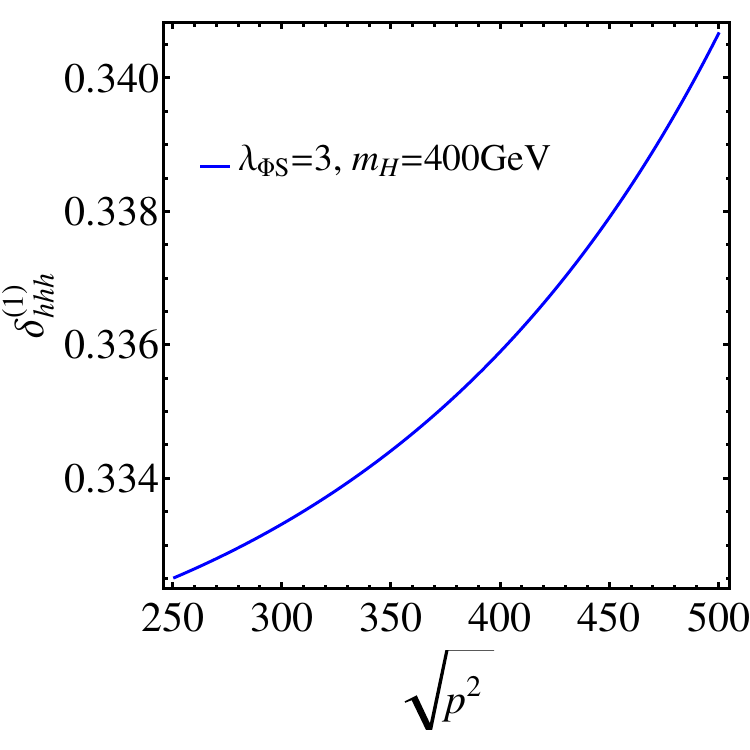}
\includegraphics[scale=0.4]{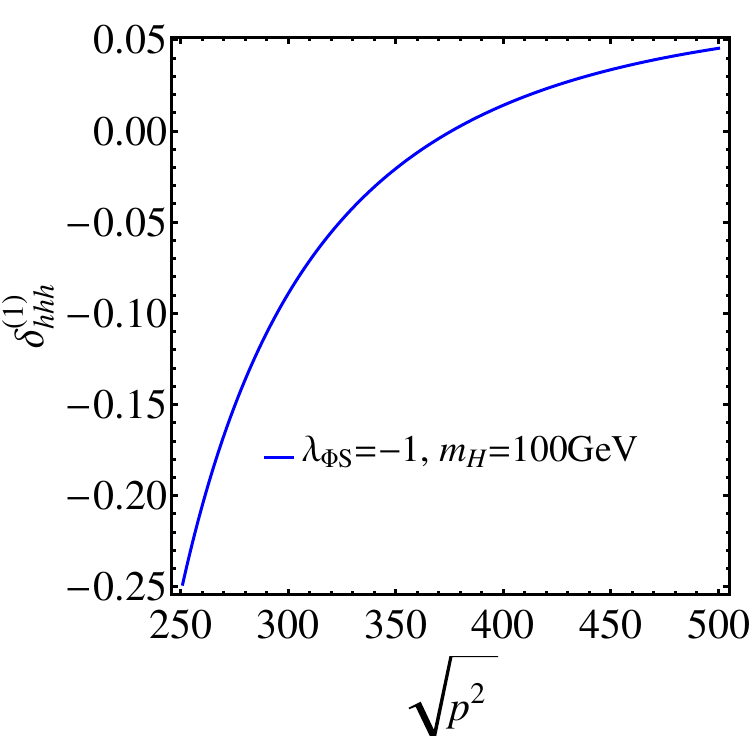}
\caption{ $\delta_{hhh}^{(1)}$ defined in Eq. \eqref{dev1} as a function of $\sqrt{p^2}~(\mathrm{GeV})$ when $(\lambda_{\Phi{S}},m_H)$ equals (1,~150GeV) (upper left), (1.5,~200GeV) (upper right), (3,~400GeV) (lower left) and (-1,~100GeV) (lower right) respectively.}
\label{fig:hhh:square}
\end{center}
\end{figure}

\begin{figure}[h]
\begin{center}
\includegraphics[scale=0.4]{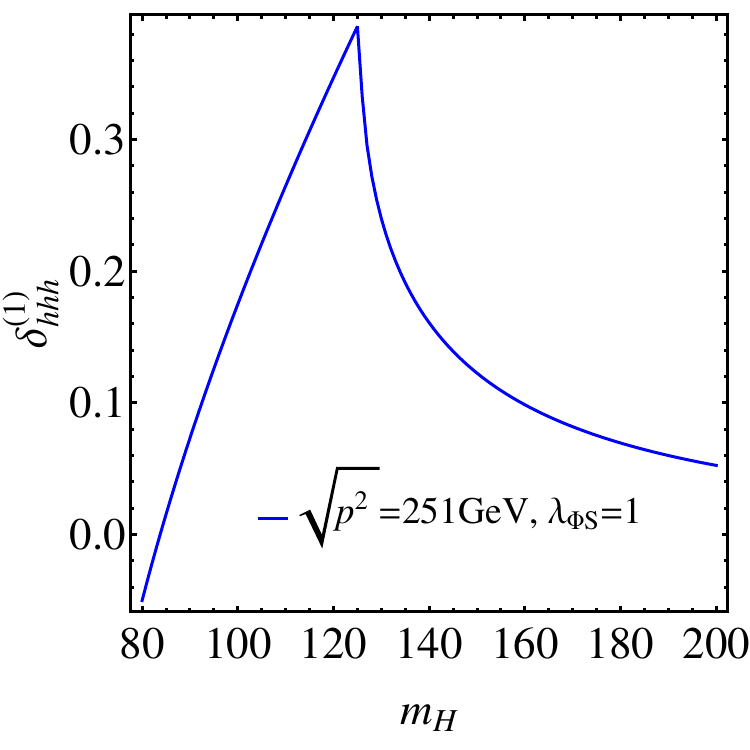}
\includegraphics[scale=0.4]{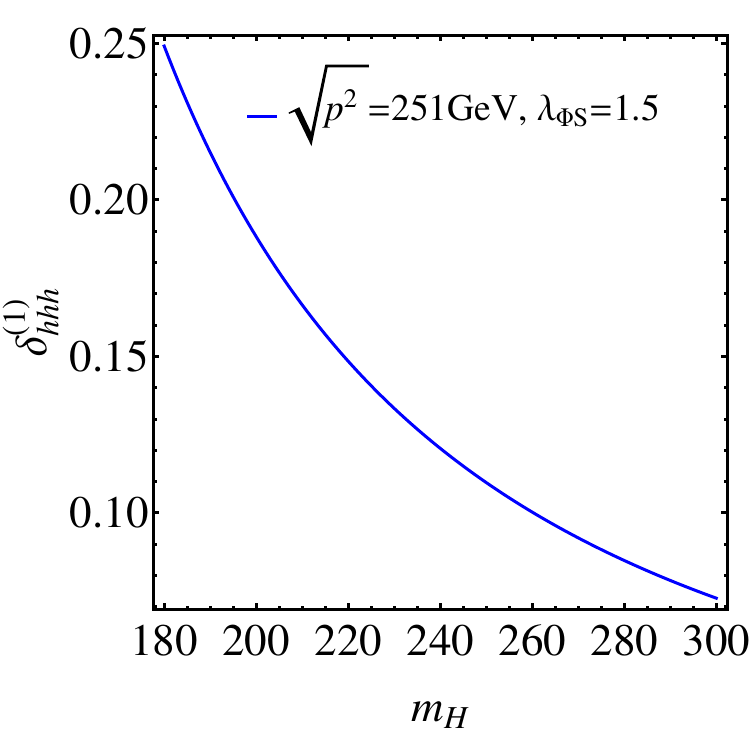}\\
\includegraphics[scale=0.4]{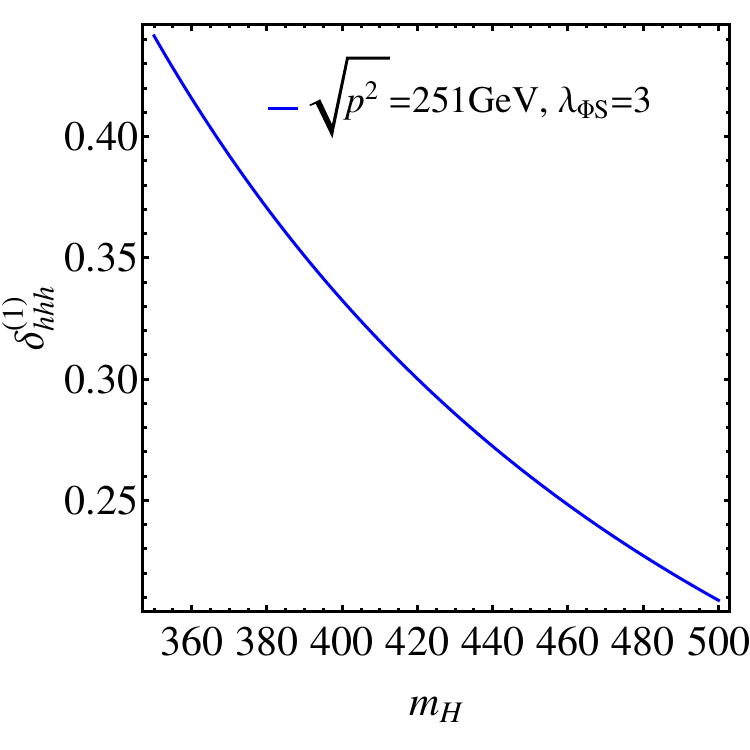}
\includegraphics[scale=0.4]{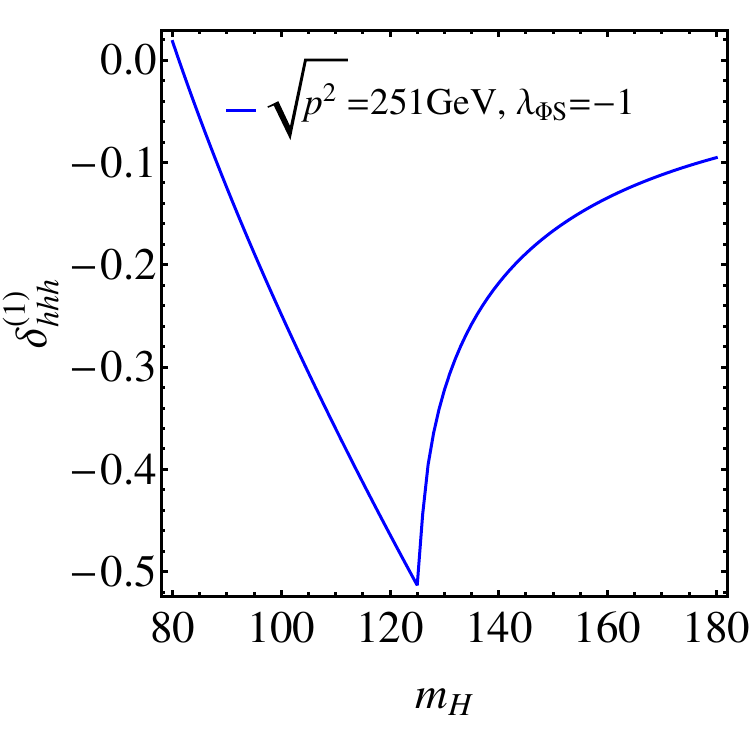}
\caption{$\delta_{hhh}^{(1)}$ defined in Eq. \eqref{dev1} as a function of $m_H~(\mathrm{GeV})$ for $\lambda_{\Phi{S}}=1$ (upper left), 1.5 (upper right), 3 (lower left) and -1 (lower right) with $\sqrt{p^2}=251\mathrm{GeV}$ respectively.}
\label{fig:hhh:mH}
\end{center}
\end{figure}

First of all, we take $\lambda_{\Phi{S}}=1,1.5,3,-1$ as reference points and study the behaviours of $\delta_{hhh}^{(1)}$. In Fig. \ref{fig:hhh:square} and Fig. \ref{fig:hhh:mH}, we show its dependence on $\sqrt{p^2}$ and $m_H$ individually. Numerical results indicate that: if $\lambda_{\Phi{S}}\sim1$, the deviation $\delta_{hhh}^{(1)}$ is positive. For $\lambda_{\Phi{S}}=1$, $\delta_{hhh}^{(1)}$ can reach $40\%$ in the vicinity of threshold when $m_H$ is light (say $m_H\in[100,140]\mathrm{GeV}$). For $\lambda_{\Phi{S}}=1.5$, $\delta_{hhh}^{(1)}$ can be $37\%$ in the vicinity of threshold when $m_H$ is about $200\mathrm{GeV}$. For $\lambda_{\Phi{S}}=3$, the deviation $\delta_{hhh}^{(1)}$ can be larger than $30\%$ when $m_H$ is near $400\mathrm{GeV}$. When $\lambda_{\Phi{S}}=-1$, the deviation $\delta_{hhh}^{(1)}$ is negative. Because $m_H$ can be light, the deviation can reach $-50\%$ in threshold enhanced parameter space. Thus precision measurements of $hhh$ coupling are sensitive to new physics.

\begin{figure}[h]
\begin{center}
\includegraphics[scale=0.4]{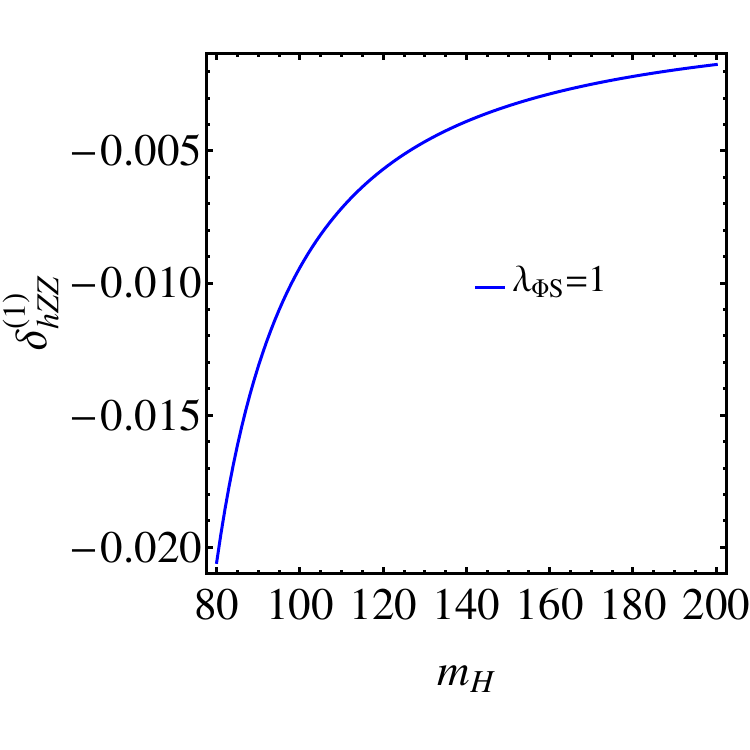}
\includegraphics[scale=0.4]{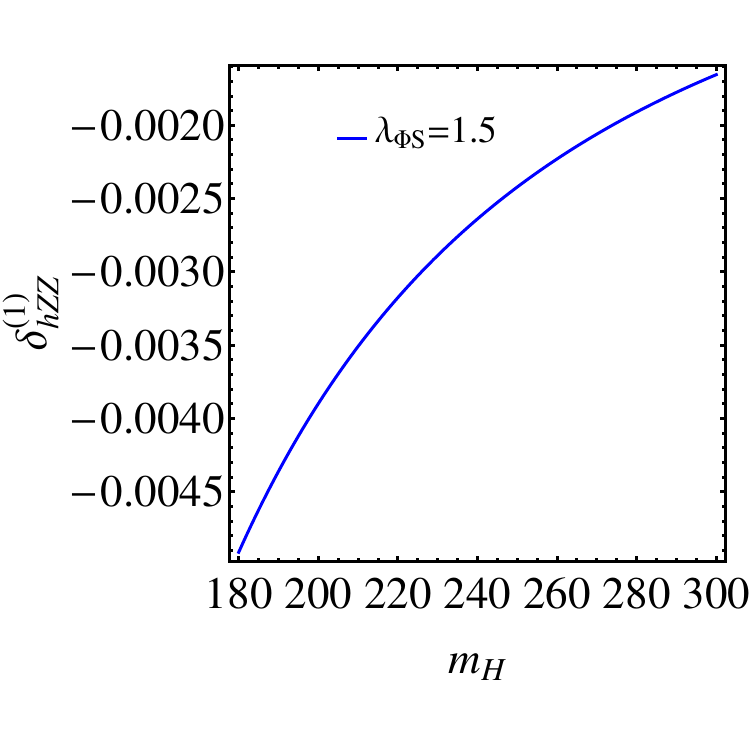}\\
\includegraphics[scale=0.4]{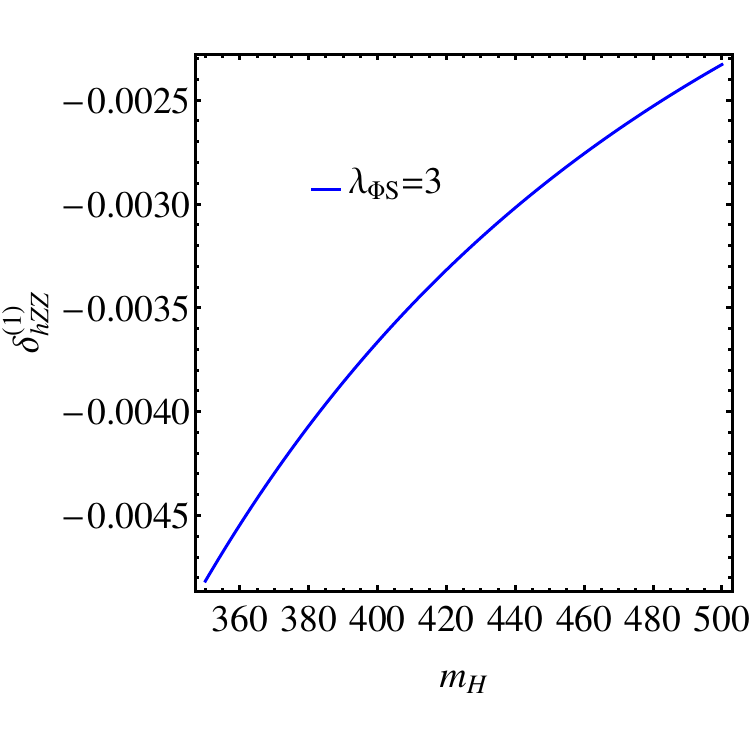}
\includegraphics[scale=0.4]{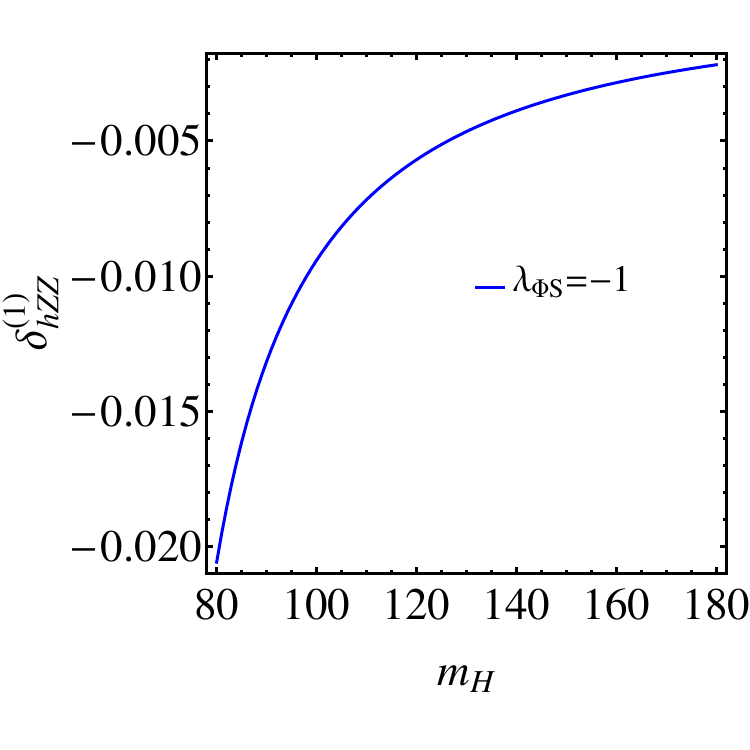}
\caption{$\delta_{hZZ}^{(1)}$ defined in Eq. \eqref{dev2} as a function of $m_H~(\mathrm{GeV})$ for $\lambda_{\Phi{S}}=1$ (upper left), 1.5 (upper right), 3 (lower left) and -1 (lower right) respectively.}
\label{fig:hZZ:mH}
\end{center}
\end{figure}

Additionally, we take $\lambda_{\Phi{S}}=1,1.5,3,-1$ as reference points and study the behaviours of $\delta_{hZZ}^{(1)}$. In Fig. \ref{fig:hZZ:mH}, we show its dependence on $m_H$. Numerical results indicate that: the deviation $\delta_{hZZ}^{(1)}$ is always negative regardless of the sign of $\lambda_{\Phi{S}}$. For $\lambda_{\Phi{S}}=1$, the deviation $\delta_{hZZ}^{(1)}$ lies between $-0.5\%\sim-2\%$ with $m_H\in[80,200]\mathrm{GeV}$. For $\lambda_{\Phi{S}}=1.5,3$, the deviation $\delta_{hZZ}^{(1)}$ ranges from $-0.2\%$ to $-0.5\%$. For $\lambda_{\Phi{S}}=-1$, it is the same as $\lambda_{\Phi{S}}=1$, and the deviation $\delta_{hZZ}^{(1)}$ can reach $-1\%$ because of relatively light $m_H$. Therefore precision measurements of $hZZ$ coupling can also be a good probe to new physics.

The sizable radiative correction for $\delta_{hhh}^{(1)}$ is mainly caused by three reasons: the magnitude of the coupling $\lambda_{\Phi{S}}$, light mass of the additional scalar and the threshold enhancement. These results are consistent with those in paper \cite{Kanemura:2016lkz}. We can find that the $\delta_{hZZ}^{(1)}$ is very small, compared to the triple $h$ coupling. Due to the high  precision measurement of $hZZ$ coupling, it can be complementary to the triple $h$ coupling to search for the BSM. By combining them together, we will detect this model well including the sign and size of $\lambda_{\Phi{S}}$.
\subsection{Present and future experimental status of the triple $h$ $\&$ $hZZ$ couplings}\label{sec:experiments}
Experimentally, the deviation of the triple $h$ coupling may be probed through double Higgs production channels at hadron colliders (such as  $gg\rightarrow{h^*}\rightarrow{hh}$). Currently, it is almost ten times larger than the SM value according to the CMS \cite{Sirunyan:2018two} and ATLAS \cite{Aad:2019uzh} collaborations. $hhh$ coupling can also be constrained from radiative corrections indirectly, but single Higgs processes \cite{Degrassi:2016wml} and EWPO \cite{Kribs:2017znd, Degrassi:2017ucl} still give the same order as those bounds from di-Higgs production. The $hhh$ coupling will be measured with $50\%$ accuracy at high luminosity LHC (HL-LHC) \cite{Dawson:2013bba, Yao:2013ika, Cao:2013cfa, Barr:2013tda, Baglio:2014aka, Cepeda:2019klc}.

The $hhh$ coupling may also be probed through $e^{+}e^{-}\rightarrow{Z^*}\rightarrow{Zh^*}\rightarrow{Zhh}$ and $e^{+}e^{-}\rightarrow{\nu_{e}\bar{\nu}_{e}W^{+*}W^{-*}}\rightarrow{\nu_{e}\bar{\nu}_{e}h^*}\rightarrow{\nu_{e}\bar{\nu}_{e}hh}$ production channels at future electron-positron colliders \cite{Abramowicz:2013tzc,Asner:2013psa,Tian:2010np}. At low energy electron-positron colliders with 240GeV or so and high luminosity, $\delta_{hhh}^{(1)}$ can also be detected indirectly \cite{McCullough:2013rea,Shen:2015pha,Cao:2014ita}. The best precision of $hhh$ coupling can be $20\%$ at future high luminosity electron-positron colliders \cite{DiVita:2017vrr}, where the deviation $\delta_{hhh}^{(1)}$ can be measured in certain parameter regions.

Up to now, global fits place restrictions on $hZZ$ coupling with ten percent uncertainty \cite{Khachatryan:2016vau, Sirunyan:2018koj}. It will be measured within $3\%$ precision at HL-LHC \cite{Cepeda:2019klc}. At future electron-positron colliders, $hZZ$ coupling precision is projected to be $0.2\%\sim0.6\%$ \cite{CEPCStudyGroup:2018ghi, Abada:2019zxq, Fujii:2017vwa, deBlas:2018mhx}.

It is of great importance to implement high precision measurements of Higgs couplings at future Higgs factories, because it may help us unveil the new physics.
\section{Conclusions}\label{sec:conclusion}
The radiative correction to the triple $h$ coupling is calculated in the minimal extension of the SM by adding a real gauge singlet scalar. In this model there are two scalars $h$ and $H$, and both of them are mixing states of the doublet and singlet. Provided that the mixing angle is set to be zero, $h$ is the pure left-over of the doublet and its behaviour is the same as that in the SM at the tree level. However the radiative corrections from the singlet $H$ can alter $h$-related couplings. $hZZ$ coupling only receives the correction from universal wave-function renormalization constant $\delta{Z_h}$ compared to the SM prediction, while $hhh$ coupling can have non-universal correction. Our numerical results show that the deviation $\delta_{hhh}^{(1)}$ is sizable. For $\lambda_{\Phi{S}}=1$, the deviation $\delta_{hhh}^{(1)}$ may lie above $20\%$ around threshold and $\delta_{hZZ}^{(1)}$ is among $-0.5\%\sim-2\%$ when $m_H$ is light. At this case, $\delta_{hZZ}^{(1)}$ will have better sensitivity to probe this model. For the parameter space with larger $\lambda_{\Phi{S}}$ (for example $\lambda_{\Phi{S}}=1.5,3$), the deviation $\delta_{hhh}^{(1)}$ grows faster than $\delta_{hZZ}^{(1)}$. $\delta_{hhh}^{(1)}$ can be more than $30\%$, while $\delta_{hZZ}^{(1)}$ is merely $-0.2\%\sim-0.5\%$. For $\lambda_{\Phi{S}}=-1$, $m_H$ can also be light. The deviation $\delta_{hhh}^{(1)}$ is negative at this moment, and it can reach $-50\%$ in threshold enhanced region. At the same time, the deviation $\delta_{hZZ}^{(1)}$ is at $-1\%$ level. The sizable radiative correction is mainly caused by three reasons: order one coupling $\lambda_{\Phi{S}}$, light mass of the additional scalar and the threshold enhancement. When searching for new physics, the model parameter space will be detected well at future high luminosity electron-positron colliders if you combine the deviation $\delta_{hhh}^{(1)}$ with $\delta_{hZZ}^{(1)}$.
\section*{Acknowledgements}
We would like to thank Kei Yagyu, Gang Li, Yang Li, Chen Zhang and Chen Shen for helpful discussions. We also thank Qing-Hong Cao, Qiang Li, Li Lin Yang, Zhao Li, Jiang-Hao Yu and Hao Zhang for pointing out some problems in our previous study. This work was supported in part by the Natural Science Foundation of China (Grants No. 11135003, No. 11635001 and No. 11375014).
\bibliographystyle{h-physrev}
\bibliography{reference}

\section*{Appendix}
\begin{appendices}
\section{Related Feynman rules}\label{app:A}
\begin{center}
\includegraphics[scale=0.5]{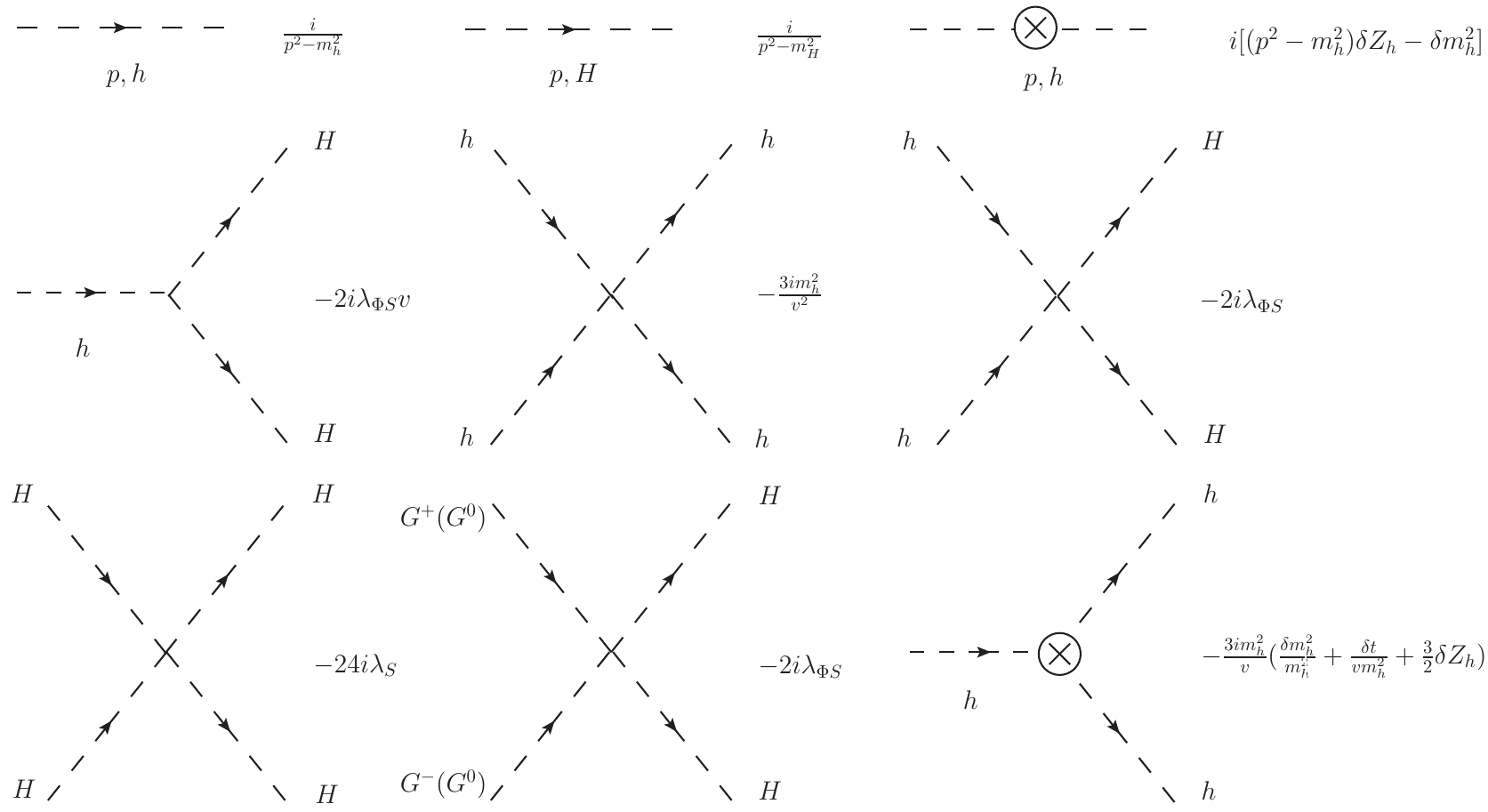}
\end{center}
\section{Perturbative unitarity constraints in the SM limit}
\label{app:unitarity}
In the SM, the perturbative unitarity constraints have been studied in the article \cite{Lee:1977eg}. For the HSM, we can write down the $7\times7$ two-to-two scattering matrix in the SM limit similarly. In the basis $W_L^+W_L^-,\frac{1}{\sqrt{2}}Z_LZ_L,hZ_L,\frac{1}{\sqrt{2}}hh,\frac{1}{\sqrt{2}}HH,hH,HZ_L$, the explicit form is shown in the following:
\begin{equation}
a_0=-\frac{1}{16\pi}\left[
\begin{array}{ccccccc}
4\lambda_{\Phi}&\sqrt{2}\lambda_{\Phi}&0&\sqrt{2}\lambda_{\Phi}&\sqrt{2}\lambda_{\Phi{S}}&0&0\\
\sqrt{2}\lambda_{\Phi}&3\lambda_{\Phi}&0&\lambda_{\Phi}&\lambda_{\Phi{S}}&0&0\\
0&0&2\lambda_{\Phi}&0&0&0&0\\
\sqrt{2}\lambda_{\Phi}&\lambda_{\Phi}&0&3\lambda_{\Phi}&\lambda_{\Phi{S}}&0&0\\
\sqrt{2}\lambda_{\Phi{S}}&\lambda_{\Phi{S}}&0&\lambda_{\Phi{S}}&12\lambda_{S}&0&0\\
0&0&0&0&0&2\lambda_{\Phi{S}}&0\\
0&0&0&0&0&0&0
\end{array}
\right].
\end{equation}
We can easily get three eigenvalues: $0,-\frac{1}{8\pi}\lambda_{\Phi},-\frac{1}{8\pi}\lambda_{\Phi{S}}$. Then, this matrix is reduced into a $4\times4$ matrix in the basis $W_L^+W_L^-,\frac{1}{\sqrt{2}}Z_LZ_L,\frac{1}{\sqrt{2}}hh,\frac{1}{\sqrt{2}}HH$:\\
\begin{equation}
a_0^{red}=-\frac{1}{16\pi}\left[
\begin{array}{cccc}
4\lambda_{\Phi}&\sqrt{2}\lambda_{\Phi}&\sqrt{2}\lambda_{\Phi}
&\sqrt{2}\lambda_{\Phi{S}}\\
\sqrt{2}\lambda_{\Phi}&3\lambda_{\Phi}&\lambda_{\Phi}&\lambda_{\Phi{S}}\\
\sqrt{2}\lambda_{\Phi}&\lambda_{\Phi}&3\lambda_{\Phi}&\lambda_{\Phi{S}}\\
\sqrt{2}\lambda_{\Phi{S}}&\lambda_{\Phi{S}}&\lambda_{\Phi{S}}&12\lambda_{S}\\
\end{array}
\right].
\end{equation}
Owing to the special structure of this matrix, we get four eigenvalues:\\
\begin{equation}
-\frac{1}{8\pi}\lambda_{\Phi},-\frac{1}{8\pi}\lambda_{\Phi},-\frac{1}{16\pi}(3\lambda_{\Phi}+6\lambda_S\pm\sqrt{(3\lambda_{\Phi}-6\lambda_S)^2+4\lambda_{\Phi{S}}^2}).
\end{equation}
It is the same as that in Ref. \cite{Kanemura:2015fra}.
\section{Tadpole and self-energy of the SM Higgs from the additional scalar}
\label{app:ren}
Tadpole of the SM Higgs and the renormalization constant $\delta{t}$:
\begin{center}
\includegraphics[scale=0.4]{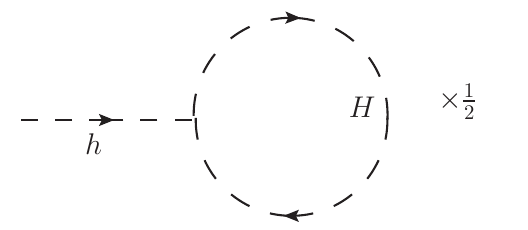}
\end{center}
\begin{equation}
iT_h=\frac{i\lambda_{\Phi{S}}v}{16\pi^2}A_0(m_H^2)~,~\delta{t}=-T_h=-\frac{\lambda_{\Phi{S}}v}{16\pi^2}A_0(m_H^2).
\end{equation}
\indent{Self-energy} of the SM Higgs and the renormalization constants $\delta{m_h^2}~,~\delta{Z_h}$:
\begin{center}
\includegraphics[scale=0.5]{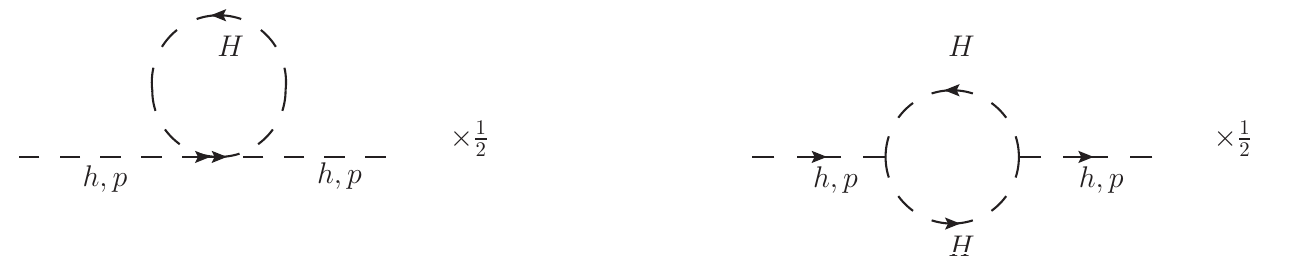}
\end{center}
\begin{equation}
\begin{aligned}
&i\Sigma_h(p^2)=\frac{i\lambda_{\Phi{S}}}{16\pi^2}A_0(m_H^2)+\frac{i\lambda_{\Phi{S}}^2v^2}{8\pi^2}B_0(p^2,m_H^2,m_H^2)\\
&\delta{m_h^2}=\mathrm{Re}\Sigma_h(m_h^2),\delta{Z_h}=-\mathrm{Re}\frac{\partial\Sigma_h(p^2)}{\partial{p^2}}|_{p^2=m_h^2}=-\frac{\lambda_{\Phi{S}}^2v^2}{8\pi^2}DB_0(m_h^2,m_H^2,m_H^2)\\
&DB_0(m_h^2,m_H^2,m_H^2)\equiv\frac{dB_0(p^2,m_H^2,m_H^2)}{dp^2}|_{p^2=m_h^2}
=\int_0^1dx\frac{x(1-x)}{m_H^2-x(1-x)m_h^2}.
\end{aligned}
\end{equation}
\section{Calculational details for the one-loop radiative corrections}
\label{app:triple}
One-loop radiative correction for the triple $h$ coupling:
\begin{center}
\includegraphics[scale=0.4]{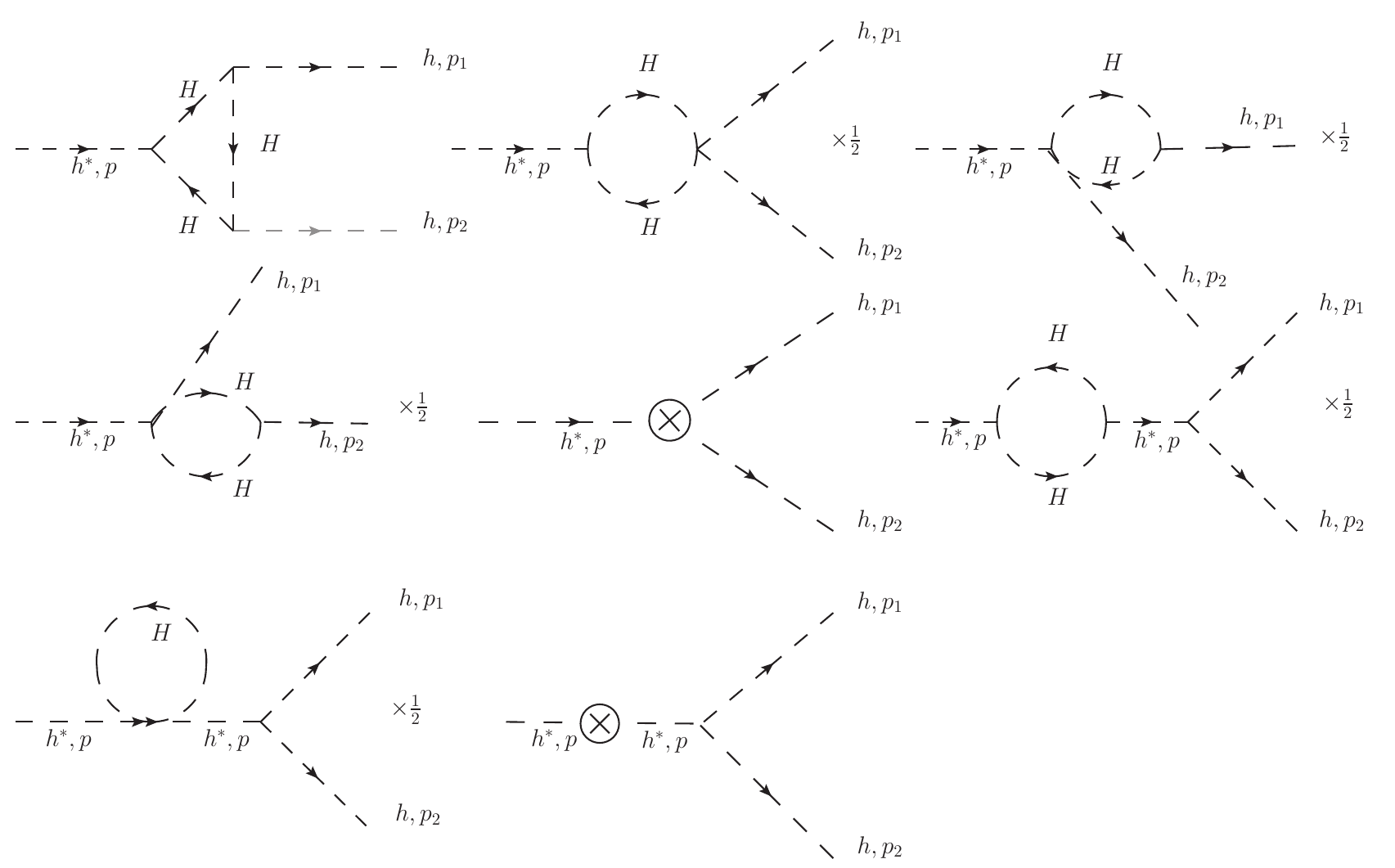}
\end{center}
Assuming the Higgs bosons with momentum $p_1,p_2$ are on shell, while the Higgs boson with momentum $p$ is off shell, that is $p_1^2=p_2^2=m_h^2,p^2\ne{m_h^2}$. We can get the following analytical expression for the triple $h$ coupling in the SM limit, which is the deviation from the SM prediction.
\begin{equation}
\begin{aligned}
&\delta_{hhh}^{(1)}\equiv\frac{\lambda_{hhh}^{(\mathrm{HSM})}-
\lambda_{hhh}^{(\mathrm{SM})}}{\lambda_{hhh}^{(\mathrm{SM},tree)}}\\
&=-\frac{\lambda_{\Phi{S}}^3v^4}{6\pi^2m_h^2}C_0(p^2,m_h^2,m_h^2,m_H^2,m_H^2,m_H^2)
+\frac{\lambda_{\Phi{S}}^2v^2}{24\pi^2m_h^2}[B_0(m_h^2,m_H^2,m_H^2)-B_0(p^2,m_H^2,m_H^2)]\\
&-\frac{\lambda_{\Phi{S}}^2v^2}{8\pi^2}\frac{B_0(p^2,m_H^2,m_H^2)-B_0(m_h^2,m_H^2,m_H^2)}{p^2-m_h^2}-\frac{\lambda_{\Phi{S}}^2v^2}{16\pi^2}\frac{\partial{B}_0(p^2,m_H^2,m_H^2)}{\partial{p}^2}|_{p^2=m_h^2}.
\end{aligned}
\end{equation}
$\lambda_{hhh}^{(\mathrm{HSM})},\lambda_{hhh}^{(\mathrm{SM})}$ are the coefficients in front of the $h^3$ vertex up to one-loop level in the HSM and SM respectively, but $\lambda_{hhh}^{(\mathrm{SM},tree)}$ is the tree level coefficient in the SM. If there is an imaginary part in $\delta_{hhh}^{(1)}$, we just extract the real part. Because the imaginary part is not observable at this order due to the interference with tree level amplitude.

Similarly, we get the one-loop radiative correction for the $hZZ$ coupling:
\begin{center}
\includegraphics[scale=0.4]{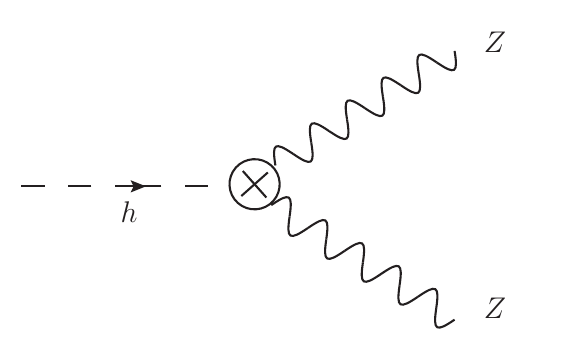}
\end{center}
\begin{equation}
\begin{aligned}
&\delta_{hZZ}^{(1)}\equiv\frac{\lambda_{hZZ}^{(\mathrm{HSM})}-
\lambda_{hZZ}^{(\mathrm{SM})}}{\lambda_{hZZ}^{(\mathrm{SM},tree)}}=\frac{\delta{Z_h}}{2}=-\frac{\lambda_{\Phi{S}}^2v^2}{16\pi^2}DB_0(m_h^2,m_H^2,m_H^2).
\end{aligned}
\end{equation}
\end{appendices}

\clearpage
\end{document}